\documentclass[%
 reprint,
superscriptaddress,
amsmath,amssymb,
prapplied
floatfix,
]{revtex4-2}

\usepackage{graphicx}
\usepackage{dcolumn}
\usepackage{bm}
\usepackage{xcolor}

\usepackage{xparse}
\usepackage{hyperref}
\usepackage[per-mode = symbol]{siunitx}
\DeclareSIUnit{\amagat}{amg}
\DeclareSIUnit{\sample}{Sa}

\usepackage[nameinlink]{cleveref}
\usepackage{makecell}
\usepackage{float}
\usepackage{tabularx}
\usepackage{array}
\usepackage{braket,bigints} 

\usepackage{bbm} 

\usepackage{lineno}

\renewcommand\bra[1]{{\langle{#1}|}}
\makeatletter
\renewcommand\ket[1]{%
\@ifnextchar\bra{\k@t{#1}\!}{\k@t{#1}}%
}
\newcommand\k@t[1]{{|{#1}\rangle}}
\makeatother

\usepackage{color}
\definecolor{mygreen}{rgb}{0,0.5,0}
\definecolor{mygrey}{rgb}{0.5,0.5,0.5}
\definecolor{myred}{rgb}{0.75,0,0}
\definecolor{myblue}{rgb}{0,0,0.75}
\definecolor{mymagenta}{cmyk}{0,1,0,0.12}
\definecolor{mycyan}{cmyk}{1,0,0,0.12}
\definecolor{myorange}{rgb}{1.,0.5,0}
\definecolor{myviolet}{rgb}{0.6,0.15,0.6}
\definecolor{mybrown}{cmyk}{0,0.50,1,0.41}

\usepackage{ulem}

\newcommand{\myexp}[1]{\exp\left[#1 \right]}

\usepackage{mathtools}
\newcommand{\Dtwo}{\ensuremath{\mathrm{D}_2}{}}
\newcommand{\Done}{\ensuremath{\mathrm{D}_1}{}}
\renewcommand{\Dtwo}{\relax\ifmmode\mathrm{D}_2\else D\textsubscript{2}{}\fi}
\renewcommand{\Done}{\relax\ifmmode\mathrm{D}_1\else D\textsubscript{1}{}\fi}

\newcommand{\mystrut}{\rule[-0pt]{0pt}{10pt}
}

\newcommand{\PDone}{\relax\ifmmode\mathrm{PD}_1\else PD\textsubscript{1}{}\fi}
\newcommand{\PDtwo}{\relax\ifmmode\mathrm{PD}_2\else PD\textsubscript{2}{}\fi}

\DeclareSIUnit{\dBm}{dBm}
\DeclareSIUnit{\torr}{Torr}

\newcommand{\ICFO}{ICFO - Institut de Ci\`encies Fot\`oniques, The Barcelona Institute of Science and Technology, 08860 Castelldefels (Barcelona), Spain}
\newcommand{\ICREA}{ICREA - Instituci\'{o} Catalana de Recerca i Estudis Avan{\c{c}}ats, 08010 Barcelona, Spain}

\begin{document}

\preprint{APS/123-QED}
\author{Christopher H. Kiehl}

\affiliation{\ICFO}

\author{Mar\'{i}a Hern\'{a}ndez Ruiz}
\affiliation{\ICFO}

\author{Cristina Sastre Jachimska} 

\affiliation{\ICFO}
\author{Morgan W. Mitchell}
\email{morgan.mitchell@icfo.eu}
\affiliation{\ICFO}
\affiliation{\ICREA}

\newcommand{\theTitle}{
Sensitivity Scaling and Limits of Cavity Enhancement in Miniaturized Optically Pumped Magnetometers
} 

\title{\theTitle}%

\date{\today}

\begin{abstract}
The sensitivity of miniaturized optically pumped magnetometers (OPMs) is limited by weak atom–light coupling, which an optical cavity can enhance. In this work, we model the photon-shot-noise-limited sensitivity of cavity-enhanced OPMs in the regime of strongly collisionally broadened optical transitions, characteristic of buffer-gas-filled miniaturized vapor cells. The cavity enhancement is benchmarked against a single-pass free-induction-decay OPM employing Faraday-rotation readout, with the probe power and detuning jointly optimized using the Cram\'{e}r--Rao lower bound as a figure of merit. For a Fabry--P\'{e}rot cavity, we compare side-of-fringe, homodyne,
Pound--Drever--Hall, and Faraday-rotation readout. All four yield an optimal
sensitivity enhancement scaling as $\alpha\sqrt{2\mathcal{F}/\pi}$, where
$\mathcal{F}$ is the cavity finesse and $0.5\leq \alpha \leq 1$ is a readout-dependent
prefactor. The enhancement is maximized at critical coupling, and we quantify its degradation away from this point. We further show that, despite spin-dependent absorption associated with the ensemble’s vector polarizability, near-critical coupling can be maintained throughout spin precession at arbitrary finesse by exceeding a derived probe-power threshold and increasing the atomic detuning with finesse. We also establish a limit to the maximum cavity enhancement set by vector light-shift noise.
\end{abstract}

\maketitle

\section{Introduction}
\label{sec:Introduction} 
Miniaturized optically pumped magnetometers (OPMs) enable smaller sensing volumes, closer standoff distances, and reduced size, weight, and power. These capabilities support improved spatial resolution in magnetic microscopy~\cite{KimAPL2017MagneticMicroscopyFluxGuides,JensenSciRep2018IsolatedAnimalHeartMCG,Hunter2026HighResolutionAtomicMagnetometerImaging},
dense sensor arrays for biomagnetic sensing and field mapping~\cite{AlemPMB2015FetalMCGMicrofabricatedOPMArray,NardelliEPJQT2020ConformalMicrofabricatedOPMArray,RasserEPJQT2025MultichannelZeroFieldOPM},
and portable magnetic sensors~\cite{SchwindtAPL2004s,mrozowski2024distributed}.
Recent advances in microfabricated vapor cells~\cite{Kitching2018ChipScaleAtomicDevices,RaghavanPRAppl2024,Wang2025MEMSAlkaliVaporCellsLaserWelding},
femtosecond-laser micromachining~\cite{LuciveroOE2022,ArtusioGlimpse2025AllDielectricVaporCells},
and integrated photonics~\cite{Hummon2018PhotonicChipLaserStabilization,Sebbag2021IntegratedNanophotonicMagnetometer}
are pushing OPM sensing volumes to the millimeter scale and below. This miniaturization, however, reduces the on-resonance optical depth of the atomic ensemble, defined by the product of atomic number density, on-resonance optical cross section, and interaction length. This quantity sets the scale of the atom–light coupling and thus the signal-to-noise ratio (SNR) of the optical readout and, ultimately, the magnetometer sensitivity.

Several factors reduce the optical depth of miniaturized OPMs. First, the shorter interaction length directly weakens it. Second, smaller cells often require higher buffer-gas pressures to suppress wall-collision-induced decoherence, while the resulting collisional broadening reduces the resonant optical cross section. The atomic density may also be constrained. Potassium magnetometers used for high-accuracy Earth-field measurements, for example, are often operated at reduced density to limit spin-exchange broadening~\cite{Pulz1999TandemMagnetometer,Aleksandrov2009ModernRadioOpticalMethods}. In close-proximity biomagnetic sensing, the vapor temperature, and therefore the atomic density, may need to remain near room temperature to avoid excessive heat transfer to the subject~\cite{barry2016optical,labyt2019magnetoencephalography}. The maximum operating temperature may also be limited by long-term vapor-cell aging arising from alkali permeation into the cell windows~\cite{KarlenOE2017, kim2025aging}. Together, these constraints motivate methods for enhancing the optical depth of the vapor cell without increasing the atom number or sensing volume.

Several strategies can, in principle, address this problem. Multipass cells enhance the accumulated optical signal by routing the probe through the atomic vapor along deterministic trajectories multiple times before detection, thereby extending the effective interaction length~\cite{silver2005simple,Li2011LargeOpticalRotation,Sheng2013SubfemtoteslaMultipass}. In Herriott-style multipass geometries used for OPMs, however, this enhancement has so far been demonstrated only in centimeter-scale cells, since miniaturizing the multipass cell itself to the millimeter scale and below imposes strict alignment and fabrication requirements~\cite{silver2005simple}. Moreover, accommodating many nonoverlapping passes within a reduced transverse area requires smaller beam spots, which can increase sensitivity to coherence loss from atomic diffusion through the focused multipass beams~\cite{Li2011LargeOpticalRotation,Sheng2013SubfemtoteslaMultipass}.
Squeezed-light probing offers another route to improved readout sensitivity~\cite{WolfgrammPRL2010, HorromPRA2012, TroullinouPRL2021, TroullinouPRL2023, SierantARX2026a}, but at the cost of substantial additional experimental complexity.

Optical cavities provide a route to enhancing the optical depth that is naturally compatible with compact sensor geometries, even
down to wavelength-scale resonators
\cite{Hunger2010FiberFabryPerot}. By resonantly recirculating the probe field,
a Fabry--P\'erot cavity increases the accumulated interaction length, giving an
effective optical-depth enhancement of order $2\mathcal{F}/\pi$~\cite{ye1998ultrasensitive}, where
$\mathcal{F}$ is the cavity finesse. This principle has been used both to enhance absorption from weak optical transitions, as in molecular gas spectroscopy~\cite{ye1998ultrasensitive} and NV-center magnetometry~\cite{ChatzidrososPRA2017}, and to increase dispersive atom--light coupling in cold-atom optical clocks~\cite{Bloom2014OpticalLatticeClock,Vallet2017NoiseImmuneCavity}, quantum non-demolition measurements using homodyne~\cite{HostenN2016,CoxPRL2016}, heterodyne~\cite{BohnetNPhot2014}, or Faraday-rotation~\cite{VasilakisNPhys2015} readout, and the recent generation of non-classical light from a thermal atomic beam~\cite{larsen2025chipscale}.

To date, cavity-enhanced OPM studies have focused primarily on optical-rotation readout, including paramagnetic Faraday rotation~\cite{MazzinghiOE2021,VasilakisNPhys2015} and nonlinear magneto-optical rotation (NMOR)~\cite{CrepazSR2015}. For paramagnetic Faraday rotation, experiments with both unpolarized~\cite{MazzinghiOE2021} and spin-polarized~\cite{VasilakisNPhys2015} ensembles, supported by theoretical analyses~\cite{Ling1994FaradayEtalon,Sycz2010ResonantFaraday,MazzinghiOE2021}, have shown that the rotation signal enhancement scales linearly with cavity finesse. At high finesse, however, atomic circular birefringence splits the cavity resonances of the two circular polarizations, thereby limiting the achievable rotation enhancement. A recent Pound--Drever--Hall (PDH) OPM based on a microfabricated vapor cell circumvented this limitation by probing a single circular-polarization eigenmode and detecting its linear dispersive phase shift~\cite{HernandezPRAppl2024,HernandezRuiz2026MagnetotacticPDH}. Despite these advances, there is still no general framework that predicts the attainable sensitivity enhancement or its dependence on cavity finesse, coupling condition, and readout scheme.

Here, we develop such a framework for cavity-enhanced OPMs with strongly collisionally broadened optical transitions, a regime characteristic of miniaturized vapor cells employing high buffer-gas pressures. We analyze a Fabry--P\'{e}rot cavity, a simple resonant geometry compatible with compact sensors, and compare four cavity-enhanced strategies for detecting the dispersive optical response arising from the vector polarizability of the atomic ensemble: side-of-fringe, homodyne, Pound--Drever--Hall, and Faraday-rotation readout. We benchmark the sensitivity enhancement against a single-pass FID OPM with Faraday-rotation readout, jointly optimizing the single-FID measurement time, probe power, and detuning to trade off signal-to-noise ratio against light-induced decoherence. The figure of merit is the photon-shot-noise-limited magnetic-field sensitivity obtained from the Cram\'{e}r--Rao lower bound (CRLB) on the variance of the estimated FID frequency. All four cavity readout schemes yield an optimal sensitivity enhancement scaling as $\alpha\sqrt{2\mathcal{F}/\pi}$, where $0.5 \leq \alpha \leq 1$ is a readout-dependent prefactor. For reflection-based readout, this optimum occurs at critical coupling, where the input-mirror transmission matches the round-trip internal loss. For transmission readout, we derive the corresponding effective critical-coupling condition. In both cases, we quantify how the enhancement degrades away from the optimum. We also show that spin-dependent absorption remains compatible with near-critical coupling at any finesse during spin precession, provided the probe power exceeds a threshold we derive and the atomic detuning is increased with finesse. Finally, we derive a limit on the maximum enhancement imposed by vector light-shift noise, which is amplified by the intracavity power.

The remainder of this article is organized as follows. In Sec.~\ref{sec:SinglePass}, we derive the relevant single-pass observables and determine the optimal sensitivity of a single-pass FID OPM. In Sec.~\ref{sec:CavityModel}, we introduce the cavity model. In Sec.~\ref{sec:cavityEnhanceSens}, we analyze the cavity-enhanced sensitivity for several detection schemes. In Sec.~\ref{sec:CritCoupOpt}, we provide a physical interpretation of the critical-coupling optimum. In Sec.~\ref{sec:cavityLimits}, we examine the practical limits on the ideal cavity enhancement set by spin-dependent absorption and vector light-shift noise. In Sec.~\ref{sec:multipass}, we compare the sensitivity enhancement of an optical cavity with that of an ideal multipass cell. We conclude in Sec.~\ref{sec:conclusion}.

\section{Optimized Single-Pass Sensitivity}
\label{sec:SinglePass}

\begin{figure*}[t]
    \centering
    \includegraphics[scale=0.47]{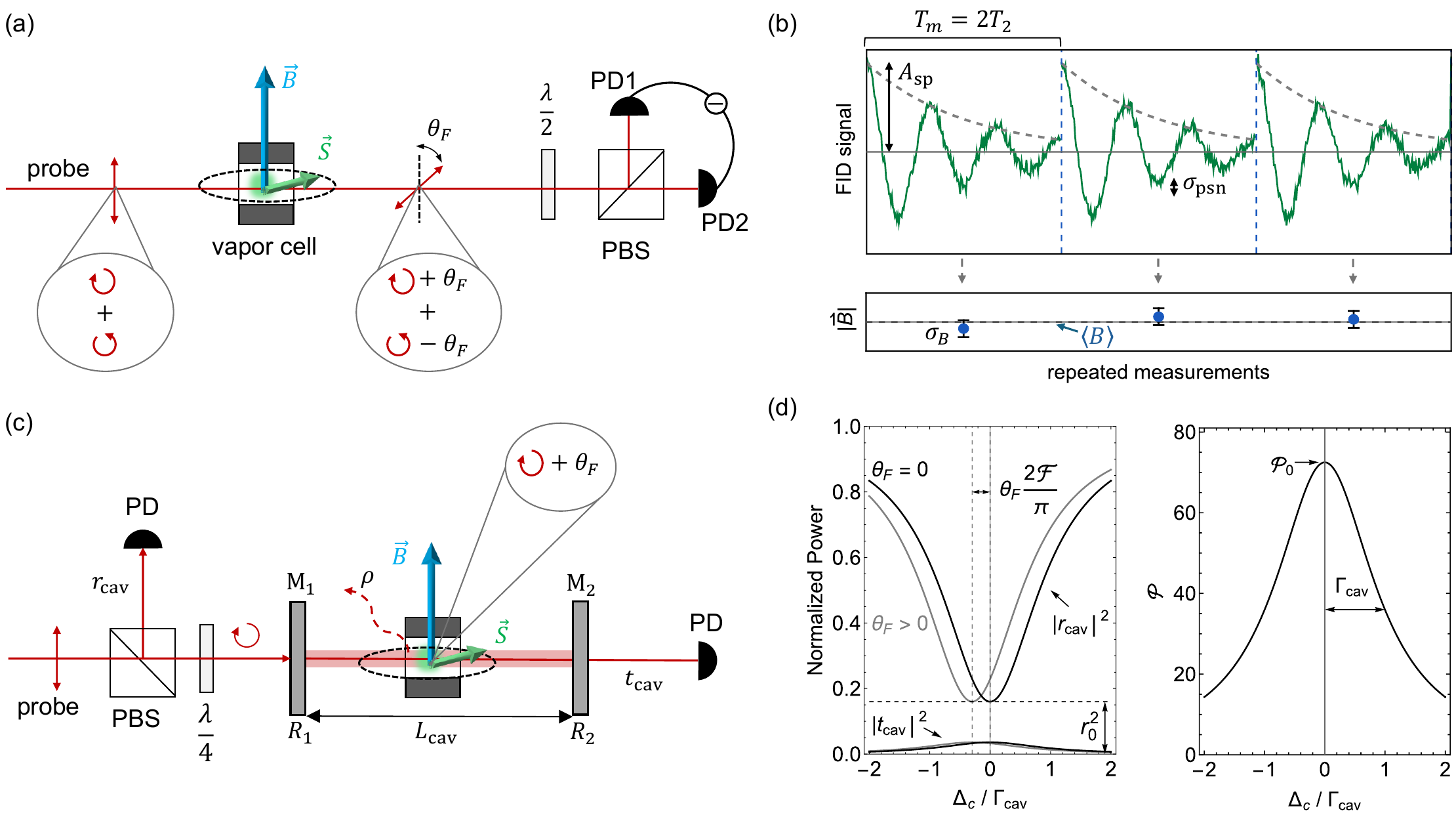}
    \caption{\textbf{Single-pass and cavity-enhanced OPM configurations.}
(a)~Single-pass geometry. An optically pumped spin polarization
$\langle\vec{S}\rangle$ precesses about the magnetic field $\vec{B}$ and is
probed by a far-detuned linearly polarized beam. The atomic circular
birefringence produces opposite phase shifts $\phi_{a}^{\pm}=\mp\theta_{F}$ on the two circular
polarization components, giving a Faraday rotation $\theta_{F}$ detected with a
balanced polarimeter.
(b)~Representative exponentially decaying free-induction-decay (FID) signal and the field estimates obtained from repeated measurements.
The relevant parameters for the Cram\'{e}r--Rao lower bound are the initial
signal amplitude $A_{\mathrm{sp}}$, the photon-shot-noise standard deviation
$\sigma_{\mathrm{psn}}$, and the measurement duration $T_m$. For white readout
noise, the optimal window is $T_m=2T_2$. Repeated FID measurements yield field
estimates with mean $\langle B\rangle$ and CRLB-limited standard deviation
$\sigma_B$, given by Eq.~\eqref{eq:field_variance}.
(c)~Fabry--P\'{e}rot cavity geometry. The vapor cell is placed inside a cavity with mirror power reflectivities $R_1=r_1^2$ and $R_2=r_2^2$. Round-trip internal losses arising, for example, from the cell windows and residual atomic absorption are modeled by the field survival factor $\rho$ defined in Eq.~\eqref{eq:round_trip_map}. Circularly polarized light is an eigenmode of the atom-coupled cavity
and acquires a phase shift $\phi_a^{\pm}=\mp\theta_F$ on each pass through the vapor cell. The
atomic phase shift can be read out in reflection or transmission through the complex
cavity amplitude coefficients $r_{\mathrm{cav}}$ and $t_{\mathrm{cav}}$.
(d)~Normalized reflected and transmitted powers, $|r_{\mathrm{cav}}|^2$ and
$|t_{\mathrm{cav}}|^2$ (left), and intracavity power enhancement
$\mathcal{P}$ relative to the single-pass case (right), plotted versus probe
detuning from the unloaded cavity resonance $\Delta_c$ in units of the cavity HWHM
linewidth $\Gamma_{\mathrm{cav}}$. Parameters are chosen for illustration:
$r_1^2=0.99$, $r_2^2=0.999$, and $r_0=-0.4$, corresponding to
$\rho\simeq0.989$. The on-resonance reflection coefficient $r_0$, defined in Eq.~\eqref{eq:r0}, determines the cavity-coupling regime, while $\mathcal{P}_0$ denotes the on-resonance intracavity power enhancement.}
    \label{fig:Figure1SPCavity}
\end{figure*}

We first establish an optimized single-pass sensitivity benchmark against which
the cavity-enhanced configurations will be compared. We use a
free-induction-decay (FID) magnetometer as a representative high-sensitivity OPM
configuration~\cite{Hunter2018FreeInductionDecayMagnetometer,Neufeld2026AtomicFreeSpinPrecession}. As shown in
Fig.~\ref{fig:Figure1SPCavity}(a), an atomic ensemble is initially
spin-polarized along the probe direction and then allowed to precess about an
applied magnetic field $\vec{B}$ at the Larmor frequency
$\omega_L=\gamma |\vec{B}|$, where $\gamma$ is the gyromagnetic ratio. Here
$\langle\vec{S}\rangle$ denotes the ensemble-averaged electron spin and
$\langle S_z\rangle$ its component along the probe axis $z$. This precession
modulates the optical response of the medium, and a far-detuned, linearly
polarized probe measures the resulting Faraday rotation,
$\theta_F\propto\langle S_z\rangle$, with a balanced polarimeter. We assume an
exponentially decaying FID signal with coherence time $T_2$ and white
photon-shot-noise-limited (PSN-limited) readout. Below, we derive the atomic
response, single-pass Faraday signal, and PSN-limited noise floor used for both
the single-pass benchmark and the cavity calculations that follow. We then use these quantities within the Cram\'{e}r--Rao lower bound
(CRLB) to obtain the optimized magnetic-field sensitivity.
\subsection{Atomic Signal and PSN-Limited Noise Floor}
We begin by deriving the atomic response that determines both the Faraday signal
and the optical absorption. The dispersive phase shifts and absorption
experienced by the probe are described by a complex transmission coefficient
$t_a^{\pm}$ acting on its circular polarization components
$E_{\mathrm{in}}^{\pm}$,
\begin{equation}
    E_{\mathrm{out}}^{\pm}
    =
    t_a^{\pm} E_{\mathrm{in}}^{\pm},
\end{equation}
where
\begin{equation}
    t_a^\pm
    =
    \myexp{
        i\frac{\omega}{c}
        \left(n_\pm - 1\right)L_a}
    \equiv
    t_s t_v^{\pm}.
    \label{eq:ta}
\end{equation}

Here $\omega$ is the optical angular frequency, $n_\pm$ are the refractive
indices for the $\sigma^\pm$ components, and $L_a$ is the length of the atomic
medium. We define the polarization components of the complex electric field as
$E_k=\hat{e}_k^*\cdot \vec{E}$, with
$\hat{e}_{\pm}=(\hat{x}\pm i\hat{y})/\sqrt{2}$ and
$\hat{e}_z=\hat{z}$. The spin-independent factor $t_s$ arises from the scalar
polarizability, while the spin-dependent factor $t_v^{\pm}$ arises from the
vector polarizability. We neglect tensor-polarizability effects, which are
suppressed when the optical linewidth is large compared with the excited-state
hyperfine splittings. This condition is met in the strongly collision-broadened regime of high-buffer-gas-pressure cells, where the collision-broadened linewidth also far exceeds the natural linewidth and collisional quenching suppresses excited-state population buildup. Power broadening and saturation are therefore negligible, and we assume throughout this article that the probe remains in the linear-response regime.

Following Ref.~\cite{HernandezPRAppl2024}, in the regime where the collisional linewidth exceeds the ground-state hyperfine splitting, the indices of refraction for the D$_1$ and D$_2$ transitions of a buffer-gas-broadened alkali vapor may be written as
\begin{align}
\begin{split}
    n_\pm - 1
    &=
    \frac{\pi n_a r_e c^2 f_{\mathrm{osc}}}{\omega_a}
    \frac{-1}{\Delta_a+i\Gamma_a}
    \left(
        1 \pm \tilde{\alpha}\,\langle S_z\rangle
    \right) \\
    &\equiv
    \delta n_s \pm \delta n_v .
    \label{eq:npm}
\end{split}
\end{align}
Here $n_a$ is the atomic number density, $r_e$ is the classical electron radius,
$f_{\mathrm{osc}}$ is the oscillator strength, $\omega_a$ is the resonance
angular frequency, $\Delta_a=\omega-\omega_a$ is the optical detuning from atomic resonance,
$\Gamma_a$ is the Lorentzian half-width at half-maximum (HWHM) in angular
frequency, and $\langle S_z\rangle$ is the expectation value of the electron-spin
projection along the probe axis, with $\langle S_z\rangle=1/2$ corresponding to
full spin polarization. The coefficient $\tilde{\alpha}$ is equal to $1$ for the
$D_2$ line and $-2$ for the $D_1$ line. The quantities $\delta n_s$ and
$\delta n_v$ denote the spin-independent and spin-dependent contributions to
$n_\pm-1$, respectively. The corresponding scalar and vector transmission
coefficients are
\begin{align}
    t_s
    &=
    \myexp{i\frac{\omega}{c}\delta n_s L_a}
    ,
        \label{eq:ts}
    \\
    t_v^\pm
    &=
    \myexp{\pm i\frac{\omega}{c}\delta n_v L_a},
    \label{eq:tvpm}
\end{align}
so that $t_a^\pm=t_s t_v^\pm$.

The Faraday rotation angle is defined as half the differential phase between the
two circular polarization components,
\begin{equation}
    \theta_F
    =
    \frac{1}{2}
    \left(
        \phi_a^- - \phi_a^+
    \right),
    \label{eq:theta_F}
\end{equation}
where
\begin{equation}
    \phi_a^\pm
    =
    \arg\left(t_v^\pm\right)
\end{equation}
is the vector-polarizability-induced phase shift acquired by the
$\sigma^\pm$ component in a single pass through the atomic medium. Since
$t_v^+$ and $t_v^-$ acquire equal and opposite dispersive phases, Eq.~\eqref{eq:theta_F} reduces to
\begin{align}
\begin{split}
    \theta_F
    & =
    \phi_a^-
    =
    -\phi_a^+ \\
    &=\pi L_a n_a r_e c f_{\mathrm{osc}} \tilde{\alpha} \langle S_z \rangle\frac{\Delta_a}{\Delta_a^2+\Gamma_a^2}
    \label{eq:phi_a}
\end{split}
\end{align}

The atomic vapor also attenuates the optical power in each circular polarization
component. For an incident field in a pure circular mode, the corresponding power
transmission is
\begin{eqnarray}
    \frac{P_{\mathrm{out}}^{\pm}}{P_{\mathrm{in}}^{\pm}}
    &=&
    \left|t_a^{\pm}\right|^2
    =
    \left|t_s\right|^2
    \left|t_v^{\pm}\right|^2
    \nonumber \\ & = &
    \myexp{-n_a\sigma_a L_a}
    \myexp{ \mp
        \frac{2\Gamma_a}{\Delta_a}
        \theta_F},
    \label{eq:atomic_power_transmission}
\end{eqnarray}
where $P_{\mathrm{in}}^{\pm}$ and $P_{\mathrm{out}}^{\pm}$ are the optical
powers in the $\sigma^\pm$ components before and after propagation through the
atomic vapor, respectively. Here $\sigma_a$ is the absorption cross section for the unpolarized atomic ensemble derived from scalar polarizability as 
\begin{equation}
\sigma_a=\frac{2\omega}{n_a c}\mathrm{Im}\big [ \delta n_s\big]=\frac{\sigma_0 \Gamma_a^2}{\Delta_a^2+\Gamma_a^2}
\label{eq:absorbcross}
\end{equation}
with the resonant absorption cross section given by 
\begin{equation}
\sigma_0=\frac{2\pi r_e c f_{\mathrm{osc}}}{\Gamma_a}.
\label{eq:sigma0}
\end{equation}

In the weak-absorption limit, the FID signal amplitude $A_{\mathrm{sp}}$,
given by the differential optical power between the two output ports of the
polarimeter PBS, is
\begin{equation}
    A_{\mathrm{sp}}
    =
    P_{\mathrm{in}}\sin\left(2\theta_F\right)
    \approx
    2P_{\mathrm{in}}\theta_F .
    \label{eq:amp_sp}
\end{equation}
Here $P_{\mathrm{in}}$ is the optical power incident on the vapor cell. The
small-angle approximation $\theta_F\ll 1$ is assumed throughout, allowing direct
comparison with the cavity-enhanced cases treated below. In principle, this
regime can always be reached by detuning the probe sufficiently far from
resonance and increasing $P_{\mathrm{in}}$ correspondingly, without degrading the
PSN-limited magnetometer sensitivity, as shown below. We also neglect non-atomic optical losses, e.g., losses in the propagation and
detection path. Non-atomic optical losses are
included in the cavity model of Sec.~\ref{sec:CavityModel}, where
they play a central role in determining the cavity buildup and extraction
efficiency. 

The photon shot-noise floor of the polarimeter is characterized by the amplitude
spectral density
\begin{equation}
    \mathcal{A}_{\mathrm{psn,sp}}
    =
    \frac{\sigma_{\mathrm{psn}}}{\sqrt{f_{\mathrm{bw,acq}}}}
    =
    \sqrt{
        2 P_{\mathrm{in}}\, \hbar\omega
    },
    \label{eq:psn_sp}
\end{equation}
with units of $\mathrm{W}/\sqrt{\mathrm{Hz}}$. Here
$\sigma_{\mathrm{psn}}$ is the standard deviation of the shot noise within the acquisition bandwidth $f_{\mathrm{bw,acq}}$. For a sample rate $f_s$,
we take $f_{\mathrm{bw,acq}}=f_s/2$. Taking the ratio of the signal amplitude
$A_{\mathrm{sp}}$ to the spectral noise floor gives the signal-to-noise ratio
in a $1~\mathrm{Hz}$ bandwidth,
\begin{equation}
    \mathrm{SNR}_{\mathrm{sp}}
    =
    \frac{A_{\mathrm{sp}}}{\mathcal{A}_{\mathrm{psn,sp}}}
    =
    \theta_F
    \sqrt{
        \frac{2 P_{\mathrm{in}}}{\hbar\omega}
    } .
    \label{eq:snr_sp}
\end{equation}
Throughout this article, SNR denotes the signal amplitude divided by the
PSN-limited noise standard deviation in a $1~\mathrm{Hz}$ bandwidth.

\subsection{Cram\'{e}r--Rao-Limited Sensitivity}

Having established the signal and noise contributions, we now derive the magnetic-field sensitivity of the FID measurement. The Cram\'{e}r--Rao lower bound
(CRLB) sets the minimum variance achievable by any unbiased estimator of the angular frequency of a discretely sampled, exponentially decaying sinusoid~\cite{Hunter2018FreeInductionDecayMagnetometer,GemmelEPJD2010}

\begin{equation}
    \sigma_\omega^2
    \geq
    \frac{12 C_{\mathrm{d}}}{\mathrm{SNR}_{\mathrm{sp}}^2 T_m^3},
    \label{eq:crlb_omega}
\end{equation}
where $T_m$ is the total measurement time and $ C_{\mathrm{d}}$ is a correction factor that
accounts for discrete sampling and the exponential decay of the FID signal.
Defining the total number of samples as $N=T_m f_s$, where $f_s$ is the sampling
frequency, this correction factor is
\begin{equation}
    C_{\mathrm{d}}
    =
    \frac{N^3}{12}
    \frac{
        \left(1-z^2\right)^3
        \left(1-z^{2N}\right)
    }{
        z^2\left(1-z^{2N}\right)^2
        -
        N^2 z^{2N}\left(1-z^2\right)^2
    },
    \label{eq:crlb_correction_factor}
\end{equation}
with
\begin{equation}
    z
    =
    \myexp{
        -1/(f_s T_2)
    },
    \label{eq:z_factor}
\end{equation}
where $T_2$ is the spin-coherence time. The frequency variance is converted to a
magnetic-field variance according to
\begin{equation}
    \sigma_B^2
    =
    \frac{\sigma_\omega^2}{\gamma^2},
    \label{eq:field_variance}
\end{equation}
where $\gamma$ is the gyromagnetic ratio.

For a sequence of FID measurements, each lasting $T_m$ and with zero dead time, the  magnetometer bandwidth (defined as the Nyquist frequency) is
$f_{\mathrm{bw,opm}}=1/(2T_m)$. The magnetic-field sensitivity, defined as
$\mathcal{A}_B=\sigma_B/\sqrt{f_{\mathrm{bw,opm}}}$, is therefore bounded by
\begin{equation}
    \mathcal{A}_B
    \geq
    \frac{\sqrt{24 C_{\mathrm{d}}}}{
        \gamma\,\mathrm{SNR}_{\mathrm{sp}}\,T_m
    }.
    \label{eq:sensitivity_tm}
\end{equation}
The measurement time that minimizes $\mathcal{A}_B$ is $T_m\approx2T_2$. For sampling
frequencies satisfying $f_s>1/T_2$, the correction factor is bounded within
$5<C_{\mathrm{d}}<8$ and may be treated as approximately constant. The resulting optimal
sensitivity is
\begin{equation}
    \mathcal{A}_B
    \geq
    \frac{\sqrt{6 C_{\mathrm{d}}}}{
        \gamma\,\mathrm{SNR}_{\mathrm{sp}}\,T_2
    }.
    \label{eq:sensitivity_general}
\end{equation}
Thus, the optimal photon-shot-noise-limited sensitivity scales inversely with
the gyromagnetic ratio $\gamma$, the single-pass signal-to-noise ratio
$\mathrm{SNR}_{\mathrm{sp}}$, and the coherence time $T_2$.

Equation~\eqref{eq:sensitivity_general} shows that optimizing the sensitivity
requires maximizing the product $\mathrm{SNR}_{\mathrm{sp}}T_2$. These two
quantities are not independent: increasing the probe power improves the SNR but
also increases light-induced spin decoherence, thereby reducing $T_2$. To determine the optimum, we consider the far-detuned regime, $\Delta_a\gg\Gamma_a$, in which combining Eqs.~\eqref{eq:phi_a} and \eqref{eq:snr_sp} gives the single-pass SNR as
\begin{equation}
    \mathrm{SNR}_{\mathrm{sp}}
    =
    C_{\mathrm{snr}}
    \frac{\sqrt{P_{\mathrm{in}}}}{\Delta_a},
    \label{eq:snr_far_detuned}
\end{equation}
where the proportionality constant is
\begin{equation}
    C_{\mathrm{snr}}
    =
    \frac{
        n_a L_a \sigma_0 \Gamma_a
        \tilde{\alpha}\langle S_z\rangle
    }{
        \sqrt{2\hbar\omega}
    }.
    \label{eq:C_snr}
\end{equation}
The rate at which an atom scatters probe photons is given by
\begin{equation}
\Gamma_{\mathrm{ph}}=\sigma_a\phi_{\mathrm{ph}},
\end{equation}
where $\sigma_a$ is defined in Eq.~\eqref{eq:absorbcross} and $\phi_{\mathrm{ph}}=P_{\mathrm{in}}/(A_{\mathrm{eff}}\hbar\omega)$ is the photon flux with $A_{\mathrm{eff}}$ the effective cross-sectional area of the probe beam. In the far-detuned limit, the photon scattering rate scales as
$\Gamma_{\mathrm{ph}}\sim1/\Delta_a^2$, and the total transverse relaxation rate can be written as
\begin{equation}
    \Gamma_2
    =
    \frac{1}{T_2}
    =\Gamma_0+\frac{\Gamma_{\mathrm{ph}}}{q}=
    \Gamma_0
    +
    C_{\Gamma}
    \frac{P_{\mathrm{in}}}{\Delta_a^2},
    \label{eq:gamma2_light_broadening}
\end{equation}
where $\Gamma_0$ is the intrinsic, power-independent relaxation rate, $q$ is the nuclear slowing-down factor, which accounts for angular momentum stored in the nuclear spin~\cite{AppeltPRA1998,Seltzer2008}. The coefficient $C_{\Gamma}$
characterizes the light-induced contribution to the transverse relaxation rate and is given by
\begin{equation}
C_{\Gamma}
    =
    \frac{
        \sigma_0 \Gamma_a^2
    }{
        q\hbar\omega A_{\mathrm{eff}}
    }.
    \label{eq:C_gamma}
\end{equation}

Since Eq.~\eqref{eq:sensitivity_general} implies
$\mathcal{A}_B\propto \Gamma_2/\mathrm{SNR}_{\mathrm{sp}}$, the optimal detuning is
found by minimizing
\begin{equation}
    \frac{\Gamma_2}{\mathrm{SNR}_{\mathrm{sp}}}
    =
    \left(
        \Gamma_0
        +
        \frac{C_{\Gamma} P_{\mathrm{in}}}{\Delta_a^2}
    \right)
    \frac{\Delta_a}{C_{\mathrm{snr}}\sqrt{P_{\mathrm{in}}}}.
    \label{eq:minimize_detuning}
\end{equation}
This minimization gives
\begin{equation}
    \Delta_a^2=
    \frac{C_{\Gamma} P_{\mathrm{in}}}{\Gamma_0} =\frac{\sigma_0 \Gamma_a^2 P_{\mathrm{in}}}{q\hbar \omega A_{\mathrm{eff}}\Gamma_0},
    \label{eq:optimalDelta}
\end{equation}
which corresponds to the condition 
\begin{equation}
    \Gamma_2=2\Gamma_0,
\label{eq:OptimumGammaRelation}
\end{equation}
or equivalently that
the light-induced relaxation rate equals the intrinsic relaxation rate $\Gamma_0=\Gamma_{\mathrm{ph}}/q$.
Substituting this optimal detuning into Eq.~\eqref{eq:sensitivity_general}
gives a single-pass sensitivity limit that is independent of both probe power and detuning,
\begin{align}
\begin{split}
    \mathcal{A}_B
    &\geq
    \sqrt{24C_{\mathrm{d}}}\,
    \frac{\sqrt{\Gamma_0 C_{\Gamma}}}{\gamma C_{\mathrm{snr}}}
    \\&=
    \frac{4\sqrt{3C_{\mathrm{d}}}}{
        \langle S_z\rangle\tilde{\alpha}
    }
    \frac{\sqrt{\Gamma_0}}{
        \gamma n_a L_a \sqrt{q\sigma_0 A_{\mathrm{eff}}}
    }.
    \label{eq:sensitivity_sp}
\end{split}
\end{align}
In general, the intrinsic relaxation rate $\Gamma_0$ depends on both the atomic
density $n_a$ and the vapor-cell length scale.
\section{Cavity Model}
\label{sec:CavityModel}

We now place the atomic vapor inside a Fabry--P\'{e}rot cavity, as illustrated in
Fig.~\ref{fig:Figure1SPCavity}(c), and analyze how the cavity modifies the
optical readout. The cavity is formed by two mirrors with amplitude reflection
coefficients $r_1$ and $r_2$ and amplitude transmission coefficients $t_1$ and
$t_2$. The corresponding power reflectivities are $R_i=r_i^2$, and, for lossless
mirrors assumed here, $t_i^2=1-R_i$. The mirrors are separated by a length $L_{\mathrm{cav}}$,
giving a free spectral range $\nu_{\mathrm{FSR}}=c/(2L_{\mathrm{cav}})$ and a
round-trip transit time $\tau_{\mathrm{rt}}=1/\nu_{\mathrm{FSR}}$. We assume that the probe is quasi-monochromatic and spatially mode-matched to the cavity TEM$_{00}$ mode. We work in the circular-polarization basis, in which
the single-pass atomic transmission coefficient $t_a^{\pm}$ is well defined and
tensor light-shift effects are negligible in the strongly Lorentzian-broadened
limit assumed throughout.

The cavity response is governed by the complex factor acquired by the intracavity field  after one round trip. Denoting the intracavity field by $E_{\mathrm{int}}$, a single round trip gives
\begin{equation}
E_{\mathrm{int}} \rightarrow r_1 r_2 \rho \myexp{i\Phi} E_{\mathrm{int}}.
\label{eq:round_trip_map}
\end{equation}
Here $\rho$, with $0 \leq \rho \leq 1$, is the round-trip field survival factor accounting for all losses not already contained in $r_1r_2$, including window losses, scattering, and atomic absorption.  The round-trip phase is
\begin{equation}
\Phi=\Delta_{\mathrm{c}}\tau_{\mathrm{rt}}+2\phi_{\mathrm{a}},
\label{eq:round_trip_phase}
\end{equation}
where the first term is the propagation phase relative to the cavity resonance and the second arises from two single passes through the atomic ensemble. Here $\phi_{\mathrm{a}}$ is the single-pass vector-polarizability phase shift defined in Eq.~\eqref{eq:phi_a}. For notational simplicity, we suppress the $\pm$ index specifying the circular polarization of the probe. The detuning $\Delta_{\mathrm{c}}=\omega-\omega_c$ is measured from the empty-cavity resonance $\omega_c$. Phase shifts arising from the cell windows and the scalar polarizability are spin independent and produce only a constant shift of the cavity resonance, which we absorb into the definition of $\omega_c$. For now, we also take $\rho$ to be spin independent. The limits to cavity enhancement that arise when spin-dependent optical loss from the vector polarizability becomes non-negligible are examined in Sec.~\ref{sec:cavityLimits}.

We adopt a phase convention in which all mirror transmission and reflection coefficients are taken to be positive real numbers, except for reflection from the external side of a cavity mirror, for which the amplitude reflection coefficient is taken to be negative real. Under these assumptions, the amplitude reflection coefficient of the cavity, $r_{\mathrm{cav}}$, after summing the reflected field over all round trips is~\cite{HernandezPRAppl2024}
\begin{equation}
r_{\rm cav}
= \frac{-r_1 + r_2\rho\,\myexp{i(\tau_{\rm rt}\Delta_c + 2\phi_a)}}
       {1 - r_1 r_2\rho\,\myexp{i(\tau_{\rm rt}\Delta_c + 2\phi_a)}}.
\label{eq:r_cav}
\end{equation}
Near resonance ($|\tau_{\rm rt}\Delta_c + 2\phi_a| \ll 1$), this expands to first order as
\begin{equation}
r_{\rm cav}
\approx r_0 + i\,r_\phi\,(\tau_{\rm rt}\Delta_c + 2\phi_a),
\label{eq:r_cav_nr}
\end{equation}
separating the response into a real on-resonance amplitude $r_0$ and a quadrature component linear in the intracavity phase shift. The on-resonance coefficient
\begin{equation}
r_0
= \frac{r_2\rho - r_1}{1 - r_1 r_2\rho}
\label{eq:r0}
\end{equation}
determines the coupling regime: $r_0 = 0$ denotes critical coupling, for which the input-mirror transmission exactly balances the round-trip internal loss; $r_0 < 0$ denotes undercoupling; and $r_0 > 0$ denotes overcoupling. The dispersive coefficient
\begin{equation}
r_\phi
= \frac{r_2\rho\,(1 - r_1^2)}{(1 - r_1 r_2\rho)^2}
\approx (1 - R_1)\frac{\mathcal{F}^2}{\pi^2}
\approx \frac{\mathcal{P}_0}{2},
\label{eq:r_phi}
\end{equation}
relates the reflected quadrature signal to the cavity finesse $\mathcal{F}$ and on-resonance intracavity power enhancement $\mathcal{P}_0$, both of which are defined below.

Similarly, the amplitude transmission coefficient is
\begin{equation}
t_{\rm cav}
= \frac{t_1 t_2\sqrt{\rho}\,
  \myexp{i(\tau_{\rm rt}\Delta_c + 2\phi_a)/2}}
  {1 - r_1 r_2\rho\,
  \myexp{i(\tau_{\rm rt}\Delta_c + 2\phi_a)}},
\label{eq:t_cav}
\end{equation}
which near resonance becomes
\begin{equation}
t_{\rm cav}
\approx t_0 + i\,t_\phi\,(\tau_{\rm rt}\Delta_c + 2\phi_a),
\label{eq:t_cav_nr}
\end{equation}
with
\begin{align}
t_0 &= \frac{t_1 t_2\sqrt{\rho}}{1 - r_1 r_2\rho},
\label{eq:t0}
\\[4pt]
t_\phi &= \frac{t_1 t_2\sqrt{\rho}\,
  (1 + r_1 r_2\rho)}{2(1 - r_1 r_2\rho)^2}.
\label{eq:t_phi}
\end{align}

For transmission measurements, the relevant coupling regime is determined by the backward coupling factor $\tilde{r}_0$, obtained from $r_0$ by interchanging $r_1$ and $r_2$:
\begin{equation}
\tilde{r}_0
= \frac{r_1\rho-r_2}{1-r_1r_2\rho}.
\label{eq:r0_tilde}
\end{equation}
For $r_1>r_2$, the condition $\tilde{r}_0=0$ defines effective critical coupling, while $\tilde{r}_0<0$ and $\tilde{r}_0>0$ correspond to effective undercoupling and overcoupling, respectively. Choosing $r_1>r_2$ is advantageous because intracavity photons preferentially exit through the output coupler $\mathrm{M}_2$ rather than through the input mirror $\mathrm{M}_1$. If the available input power is not limited, increasing $r_1$ carries no fundamental penalty, since the weaker input coupling can be compensated by increasing $P_{\mathrm{in}}$ to maintain the same intracavity power. In practice, $r_1$ should be chosen large enough to favor transmission while still providing sufficient detected power.

To account for additional optical power at the vapor cell, we define the ratio of the intracavity power at the position of the atomic vapor to the single-pass power through the same medium as
\begin{equation}
\mathcal{P}
= \frac{P_{\rm cav}}{P_{\rm in}}
= \frac{2(1-R_1)}
  {\bigl|1 - r_1 r_2\rho\,
  \myexp{i(\tau_{\rm rt}\Delta_c + 2\phi_a)}\bigr|^2}.
\label{eq:intracavity_power}
\end{equation}
Near resonance this takes a Lorentzian form,
\begin{equation}
\mathcal{P}
\approx \frac{\mathcal{P}_0}
  {1 + (\Delta_c/\Gamma_{\rm cav})^2},
\label{eq:P_lorentzian}
\end{equation}
where the cavity HWHM linewidth in angular-frequency units,
\begin{equation}
\Gamma_{\rm cav}
= \frac{\nu_{\rm FSR}(1 - r_1 r_2\rho)}{r_1 r_2\rho},
\label{eq:gamma_cav}
\end{equation}
is identified from the denominator of Eq.~\eqref{eq:intracavity_power}, and the peak enhancement is
\begin{equation}
\mathcal{P}_0
= \frac{2(1 - R_1)}{(1 - r_1 r_2\rho)^2}
\approx 2(1 - R_1)\frac{\mathcal{F}^2}{\pi^2}.
\label{eq:P0}
\end{equation}
The cavity finesse,
\begin{equation}
\label{eq:finesse}
\mathcal{F}
= \frac{\pi\nu_{\rm FSR}}{\Gamma_{\rm cav}}
= \frac{\pi\,r_1 r_2\rho}{1 - r_1 r_2\rho}
\approx \frac{\pi}{1 - r_1 r_2\rho},
\end{equation}
is the ratio of the free spectral range to the FWHM linewidth $\delta\nu_{\rm cav} = \Gamma_{\rm cav}/\pi$; the last approximation holds for $1 - r_1 r_2\rho \ll 1$.

A round-trip phase shift \(2\phi_a\) imparted to a given circular polarization component shifts the corresponding cavity resonance by angular frequency
\begin{equation}
\label{eq:cavity_shift}
\delta\Delta_c
= -2\phi_a\nu_{\rm FSR}
= -2\theta_F\nu_{\rm FSR},
\end{equation}
where Eq.~\eqref{eq:phi_a} gives \(\phi_a = \theta_F\). The ratio of this atom-induced shift to the cavity linewidth is then
\begin{equation}
\frac{\delta\Delta_c}{\Gamma_{\rm cav}}
= -\theta_F\frac{2\mathcal{F}}{\pi}.
\label{eq:shift_ratio}
\end{equation}
In this sense, the cavity enhances the effect of circular birefringence by a factor of $2\mathcal{F}/\pi$ relative to the single-pass case. 

\section{Cavity-Enhanced Sensitivity}
\label{sec:cavityEnhanceSens}
We apply the cavity model of Sec.~\ref{sec:CavityModel} to quantify the
sensitivity enhancement of a cavity-based OPM relative to an optimized
single-pass OPM. We evaluate four schemes for detecting the cavity-induced phase
shift: side-of-fringe, homodyne, PDH, and Faraday rotation. Although these
schemes differ in their experimental complexity and susceptibility to technical
noise, at critical coupling they all yield nearly the same optimal sensitivity
enhancement,
$\alpha\sqrt{2\mathcal{F}/\pi}$,
as summarized in Table~\ref{tab:readout_schemes}, where
$0.5 \leq \alpha \leq 1$ is a readout-dependent prefactor.

As shown in Eq.~\eqref{eq:sensitivity_general}, determining the sensitivity
enhancement provided by the cavity OPM requires only quantifying how the cavity
modifies the SNR and the transverse relaxation rate $\Gamma_2$ relative to a
single-pass OPM. Thus, for each detection method, we first calculate the
signal-to-noise ratio gain $G_{m,c}$ of the cavity readout relative to the
optimized single-pass readout under otherwise identical conditions:
\begin{equation}
G_{m,c}
=
\frac{\mathrm{SNR}_{m,c}}{\mathrm{SNR}_{\mathrm{sp}}}.
\label{eq:genericGain}
\end{equation}
Here, $m$ denotes the detection method, for example, $m=\mathrm{sf}$ for
side-of-fringe detection, and $c\in\{\mathrm{r},\mathrm{t}\}$ denotes detection channel
in reflection or transmission, respectively. Analogously to the single-pass
case in Eq.~\eqref{eq:snr_sp}, the signal-to-noise ratio of the cavity readout
in a 1-Hz bandwidth is defined as
\begin{equation}
\mathrm{SNR}_{m,c}
=
\frac{A_{m,c}}{\mathcal{A}_{\mathrm{psn},m,c}}.
\end{equation}
Here, $A_{m,c}$ denotes the cavity FID signal amplitude, corresponding to the
single-pass amplitude defined in Eq.~\eqref{eq:amp_sp}, and
$\mathcal{A}_{\mathrm{psn},m,c}$ denotes the photon-shot-noise floor of the
cavity readout, corresponding to the single-pass quantity defined in
Eq.~\eqref{eq:psn_sp}.

Maintaining the optimal transverse relaxation condition
$\Gamma_2=2\Gamma_0$ (\autoref{eq:OptimumGammaRelation}, derived in Sec.~\ref{sec:SinglePass}), in the presence of
the intracavity power enhancement $\mathcal{P}_{m,c}$, which depends on the
readout method and detection channel,  requires {the change $P_{\mathrm{in}}/\Delta_a^2 \rightarrow P_{\mathrm{in}} /\Delta_a^2\mathcal{P}_{m,c}$. Here, $P_{\mathrm{in}}$ denotes the probe power in the absence of cavity mirrors, i.e., for the single-pass OPM described in Sec.~\ref{sec:SinglePass}, the optical power incident on the cell. This condition can be achieved by increasing the detuning according to
$\Delta_a\rightarrow\Delta_a\sqrt{\mathcal{P}_{m,c}}$ or by reducing the incident probe power according to
$P_{\mathrm{in}}\rightarrow P_{\mathrm{in}}/\mathcal{P}_{m,c}$.} 
{As a result, under the constraint that the optical-scattering-induced decoherence remains equal to that of the optimized single-pass OPM, the SNR gain calculated in Eq.~\eqref{eq:genericGain} is reduced by a factor of $\sqrt{\mathcal{P}_{m,c}}$.} The corresponding magnetic-sensitivity enhancement
is therefore
\begin{equation}
\mathcal{E}_{m,c}
=
\frac{G_{m,c}}{\sqrt{\mathcal{P}_{m,c}}}.
\label{eq:genericEnhancement}
\end{equation}

For each readout scheme, we optimize $\mathcal{E}_{m,c}$ with respect to the relevant cavity detuning. The resulting optimal condition may correspond to operation off resonance, as in side-of-fringe detection, or to resonant operation of the relevant optical component, for example, the carrier in PDH detection. In every case, we assume that the cavity-frequency shift induced by the atomic phase shift $\phi_a$ remains sufficiently small that the optical response stays within its linear regime. Alternatively, feedback could be used to continuously maintain the optimal detuning condition, for example, by tuning the laser frequency. In this case, the signal would be inferred from the applied feedback. 

Throughout the following subsections, numerical calculations and plots assume a cavity length $L_{\mathrm{cav}}=1~\mathrm{cm}$, corresponding to a free spectral range $\nu_{\mathrm{FSR}}=c/(2L_{\mathrm{cav}})\approx15~\mathrm{GHz}$.

\begin{table}[t]
    \caption{Comparison of the optimal cavity enhancement for the different reflection and transmission readout methods, evaluated at their respective critical-coupling conditions: $r_0=0$ in reflection and $\tilde r_0=0$ in transmission.}
    \label{tab:readout_schemes}
    \begin{ruledtabular}
        \begin{tabular}{lcc}
            Readout scheme & $\mathcal{E}_{m,\mathrm{r}}$ & $\mathcal{E}_{m,\mathrm{t}}$ \\
            \hline
            Side-of-fringe & $\sqrt{2\mathcal{F}/\pi}$ & $\sqrt{\mathcal{F}/2\pi}$ \\
            Homodyne       & $\sqrt{2\mathcal{F}/\pi}$ & $\sqrt{2\mathcal{F}/\pi}$ \\
            PDH            & $\sqrt{4\mathcal{F}/3\pi}$ & --- \\
            Faraday        &--- & $\sqrt{2\mathcal{F}/\pi}$
        \end{tabular}
    \end{ruledtabular}
\end{table}

\begin{figure*}[t]
    \centering
    \includegraphics[scale=0.68]{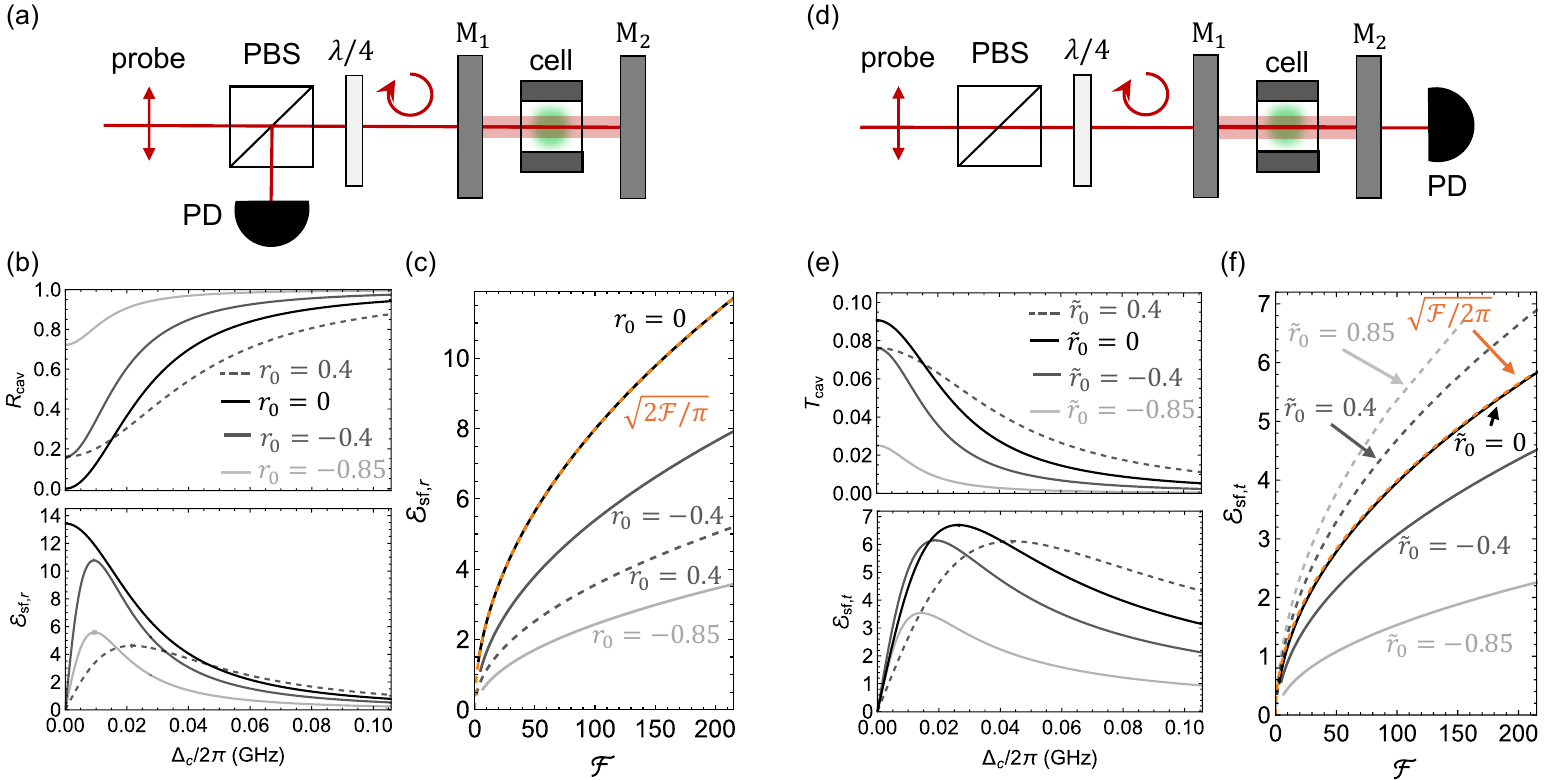}
    \caption{\textbf{Side-of-fringe sensitivity enhancement.} Throughout, we assume a cavity length $L_{\mathrm{cav}}=1~\mathrm{cm}$, corresponding to
    a free spectral range $\nu_{\mathrm{FSR}}\approx15~\mathrm{GHz}$.
    (a) Reflection-detection apparatus.
    (b) Cavity reflectivity and sensitivity enhancement versus cavity detuning for different degrees of undercoupling (solid) and overcoupling (dashed). The intracavity loss is fixed at $\rho=0.995$, $r_2^2=0.999$, and $r_1^2$ is chosen to obtain the specified value of $r_0$.
    (c) Reflection sensitivity enhancement versus finesse for different cavity couplings. For each finesse, $r_2^2=0.999$ is fixed and $\rho$ is varied, after which $r_1$ is chosen to obtain the specified value of $r_0$.
    (d) Transmission-detection apparatus.
    (e) Cavity transmission and sensitivity enhancement versus cavity detuning for different degrees of undercoupling (solid) and overcoupling (dashed). The intracavity loss is fixed at $\rho=0.995$, $r_1^2=0.999$, and $r_2^2$ is chosen to obtain the specified value of $\tilde{r}_0$.
    (f) Transmission sensitivity enhancement versus finesse for different cavity couplings. For each finesse, $r_1^2=0.999$ is fixed and $\rho$ is varied, after which $r_2$ is chosen to obtain the specified value of $\tilde{r}_0$.
    Because the same finesse can arise from different combinations of $r_2$ and $\rho$, the improved overcoupled scaling in transmission results from increasing $\rho$, corresponding to lower internal loss, while decreasing $r_2$ at fixed finesse.} 
    \label{fig:sideOfFringe}
\end{figure*}

\subsection{Side-of-Fringe Detection}

In side-of-fringe detection, a circularly polarized probe is detuned from a cavity resonance so that an atom-induced resonance shift produces a linear change in the reflected or transmitted optical power. To derive the side-of-fringe FID signal amplitude $A_{\mathrm{sf}}$, we define the unloaded normalized cavity reflection and transmission powers as
\begin{align}
R_{\mathrm{cav}}(\Delta) &\equiv \left. | r_\mathrm{cav}\mystrut|^2 \right|_{\scriptsize\begin{array}{l}\phi_a=0 \\ \Delta_c = \Delta \end{array} }
\\ T_{\mathrm{cav}}(\Delta) &\equiv \left. |t_\mathrm{cav}\mystrut |^2 \right|_{\scriptsize\begin{array}{l}\phi_a=0 \\ \Delta_c = \Delta \end{array}},
\end{align}
respectively, where $r_{\mathrm{cav}}$ and $t_{\mathrm{cav}}$ are given by
Eqs.~\eqref{eq:r_cav} and \eqref{eq:t_cav}. The corresponding reflection and
transmission slopes are defined as
\begin{align} 
s_{\mathrm{R}}(\Delta_c)&=\left. \frac{d}{d\Delta} R_{\mathrm{cav}}(\Delta)\right|_{\Delta = \Delta_c}\\ s_{\mathrm{T}}(\Delta_c)&=\left. \frac{d}{d\Delta} T_{\mathrm{cav}}(\Delta)\right|_{\Delta = \Delta_c}.
\end{align}

For reflection detection, illustrated in
Fig.~\ref{fig:sideOfFringe}(a), the FID signal amplitude is obtained by multiplying the slope of the unloaded-cavity reflection spectrum
evaluated at the unloaded detuning $\Delta_{c}$, by the atom-induced cavity shift $\delta\Delta_c$:
\begin{align}
\begin{split}
A_{\mathrm{sf,r}}
&=
\delta\Delta_c P_{\mathrm{in}}s_{\mathrm{R}}(\Delta_{c})
\\
&=
2P_{\mathrm{in}}\theta_F
\Bigg[
\nu_{\mathrm{FSR}} s_R(\Delta_{c})
\Bigg].
\label{eq:Asf}
\end{split}
\end{align}
In the second equality,
Eq.~\eqref{eq:cavity_shift} has been used to express the atom-induced cavity
shift in terms of the Faraday rotation angle $\theta_F$. The bracketed term is therefore the signal gain relative to the single-pass Faraday signal Eq.~\eqref{eq:amp_sp}. 
The reflected side-of-fringe PSN floor is reduced relative to the single-pass
value in Eq.~\eqref{eq:psn_sp} by the ratio
\begin{equation}
\frac{\mathcal{A}_{\text{psn,sf}}}{\mathcal{A}_{\text{psn,sp}}}
=
\sqrt{R_{\mathrm{cav}}(\Delta_{c})}.
\label{eq:psn_ratio_sf}
\end{equation}
Combining Eq.~\eqref{eq:Asf} and Eq.~\eqref{eq:psn_ratio_sf} gives the SNR gain as defined in Eq.~\eqref{eq:genericGain}
\begin{equation}
G_{\text{sf,r}}
=
\Bigg[
\nu_{\text{FSR}}s_{\mathrm{R}}(\Delta_{c})
\Bigg]
\frac{1}{\sqrt{R_{\mathrm{cav}}(\Delta_{c})}}.
\end{equation}
Using Eq.~\eqref{eq:genericEnhancement} to account for the effect of the
intracavity power enhancement on the transverse
relaxation rate, the sensitivity enhancement for reflection detection is
\begin{equation}
\mathcal{E}_{\mathrm{sf,r}}
=
\Bigg[
\nu_{\text{FSR}} s_{\mathrm{R}}(\Delta_{c})
\Bigg]
\frac{1}
{\sqrt{R_{\mathrm{cav}}(\Delta_{c})\mathcal{P}(\Delta_{c})}}.
\end{equation}
Here, $\mathcal{P}(\Delta_{c})$ is the intracavity power enhancement defined in
Eq.~\eqref{eq:intracavity_power} and evaluated at detuning $\Delta_{c}$ for the unloaded cavity, $\phi_a=0$. Following the same reasoning, the sensitivity enhancement for transmission detection in the configuration shown in Fig.~\ref{fig:sideOfFringe}(d) is
\begin{equation}
\mathcal{E}_{\mathrm{sf,t}}
=
\Bigg[
\nu_{\text{FSR}}s_{\mathrm{T}}(\Delta_{c})
\Bigg]
\frac{1}
{\sqrt{T_{\mathrm{cav}}(\Delta_{c})\mathcal{P}(\Delta_{c})}}.
\end{equation}

Figure~\ref{fig:sideOfFringe}(b,e) shows the normalized reflection and transmission powers, $R_{\mathrm{cav}}(\Delta_{c})$ and $T_{\mathrm{cav}}(\Delta_c)$, and their corresponding sensitivity enhancements versus cavity detuning for several coupling regimes at fixed round-trip field survival $\rho=0.995$. For reflection detection, $r_2^2=0.999$ is fixed and $r_1$ is chosen to realize the reflection coupling factor $r_0$ defined in Eq.~\eqref{eq:r0}. For transmission detection, the mirror roles are interchanged, $r_1 \leftrightarrow r_2$, and $r_2$ is chosen to realize the backward coupling factor $\tilde{r}_0$ defined in Eq.~\eqref{eq:r0_tilde}. After optimization over $\Delta_c$, the sensitivity enhancements $\mathcal{E}_{\mathrm{sf},r},\mathcal{E}_{\mathrm{sf},t}$ shown in Fig.~\ref{fig:sideOfFringe}(b,e) are maximized under the critical-coupling conditions $r_0,\tilde{r}_0=0$, respectively. 

Notably, for critically coupled reflection detection, the optimum is approached in the limit $\Delta_c\rightarrow0$. In this limit, both the reflected signal amplitude and the reflected photon-shot-noise contribution vanish, but their ratio remains finite. This result should therefore be interpreted as an ideal photon-shot-noise-limited bound rather than a practical operating point. Additional noise sources, such as electronic noise, generally shift the optimum to a finite detuning with nonzero detected power. When these contributions are small, the achievable sensitivity may nevertheless remain close to the ideal limit.

Figure~\ref{fig:sideOfFringe}(c,f) shows the detuning-optimized sensitivity
enhancement as a function of finesse. As in
Fig.~\ref{fig:sideOfFringe}(b,e), the power reflectivity of the mirror opposite
the input port is fixed at $r_i^2=0.999$. To vary the finesse, the round-trip
field survival $\rho$ is varied, while the coupling-mirror reflectivity is
chosen to realize the specified coupling condition. Numerically, the
enhancement in reflection at $r_0=0$ is found to follow
\begin{equation}
    \mathcal{E}_{\mathrm{sf,r}}^{(\mathrm{cc})}
    \simeq
    \sqrt{\frac{2\mathcal{F}}{\pi}},
\end{equation}
and in transmission at $\tilde{r}_0=0$ to follow
\begin{equation}
    \mathcal{E}_{\mathrm{sf,t}}^{(\mathrm{cc})}
    \simeq
    \sqrt{\frac{\mathcal{F}}{2\pi}}.
\end{equation}
At fixed finesse, departures from critical coupling in
reflection detection reduce $\mathcal{E}_{\mathrm{sf,r}}$, as shown in
Fig.~\ref{fig:sideOfFringe}(c). Transmission detection, shown in
Fig.~\ref{fig:sideOfFringe}(f), instead appears to favor the effective
overcoupled regime $\tilde{r}_0>0$. 

The above results provide a simple relationship between finesse and optimal sensitivity enhancement for this readout strategy. In the case of reflection, they also indicate that critical coupling is optimal, which implies an unambiguous relationship between $\rho$ and $r_1$, given that $r_2\approx 1$. 

In contrast, for transmission, the optimum is over-coupled, in the sense that $\tilde{r}_0 > 0$. It should be noted that this does not uniquely define $r_2$, because $\mathcal{F}$ depends on both $\rho$ and $r_2$.  In many contexts it might be interesting to consider fixed $\rho$, as done in Figure~\ref{fig:sideOfFringe}(b,e).

\begin{figure*}[t]
    \centering
    \includegraphics[scale=0.6]{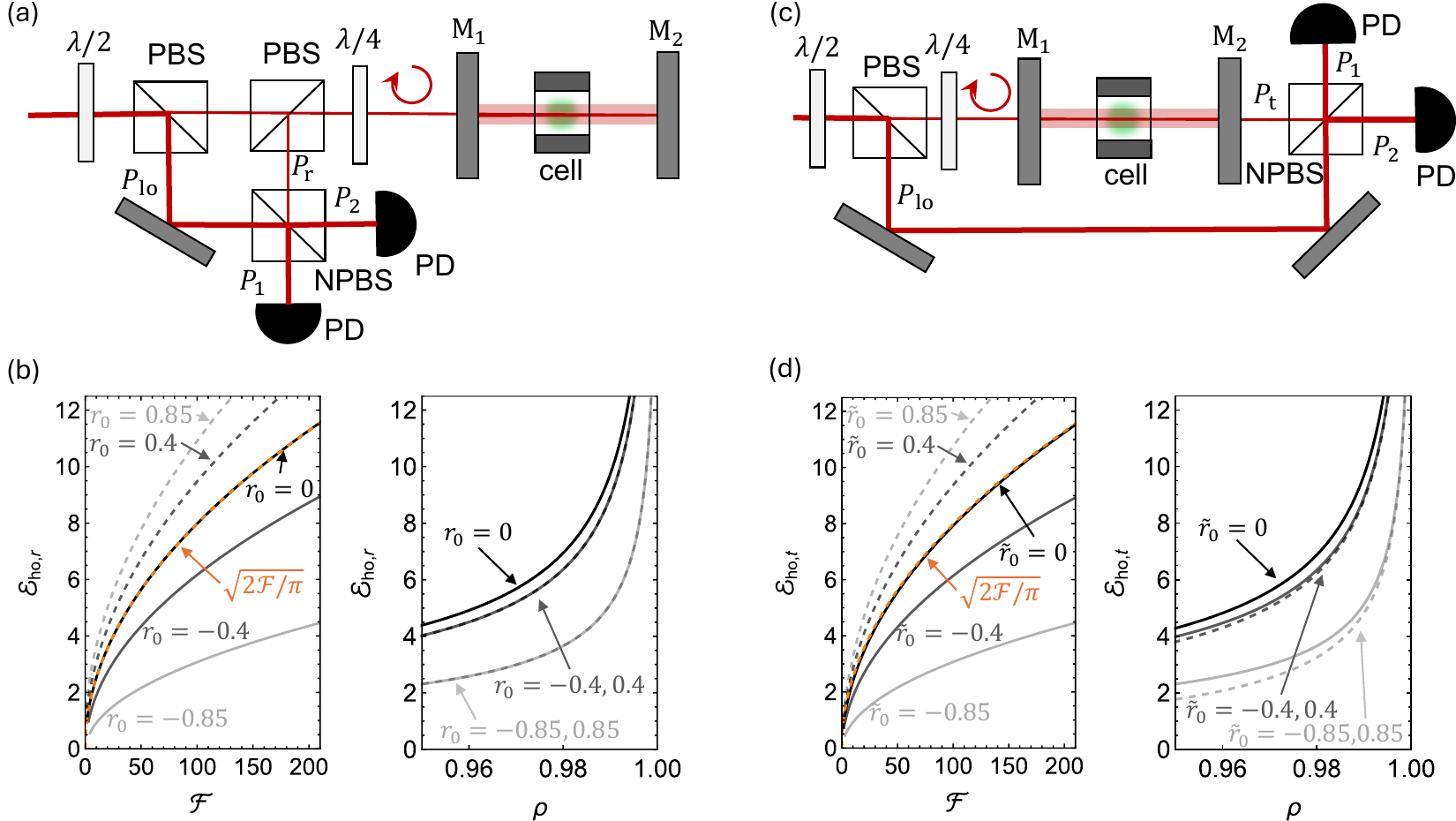}
    \caption{\textbf{Homodyne sensitivity enhancement.} Throughout, we assume a cavity length $L_{\mathrm{cav}}=1~\mathrm{cm}$, corresponding to
    a free spectral range $\nu_{\mathrm{FSR}}\approx15~\mathrm{GHz}$.
    (a) Apparatus for homodyne detection in reflection.
    (b) Optimized reflection sensitivity enhancement as a function of cavity finesse $\mathcal{F}$ (left) and internal round-trip amplitude transmission $\rho$ (right) for several coupling conditions. For each finesse, $r_2^2=0.999$ is fixed, $\rho$ is varied, and $r_1$ is chosen to obtain the specified value of $r_0$.
    (c) Apparatus for homodyne detection in transmission.
    (d) Optimized transmission sensitivity enhancement as a function of $\mathcal{F}$ (left) and $\rho$ (right) for several coupling conditions. For each finesse, $r_1^2=0.999$ is fixed, $\rho$ is varied, and $r_2$ is chosen to obtain the specified value of $\tilde{r}_0$.
    For the calculations in (b) and (d) the probe is assumed to remain on cavity resonance, $\Delta_c=0$, either through active feedback or in the small-shift limit $\phi_a\ll1$.
    At fixed $\rho$, the maximum enhancement occurs at critical coupling, corresponding to $r_0=0$ in reflection and $\tilde{r}_0=0$ in transmission.}
    \label{fig:homodyne}
\end{figure*}

\subsection{Homodyne Detection}

Homodyne detection measures the atom-induced cavity-resonance shift by interfering the reflected or transmitted cavity field with a local-oscillator field, $E_{\mathrm{lo}}$. We first consider homodyne detection in reflection using the configuration shown in Fig.~\ref{fig:homodyne}(a). From Eq.~\eqref{eq:r_cav_nr}, the reflected field on unloaded cavity resonance is
\begin{equation}
E_r=E_{\text{in}}(r_0+ir_{\phi}(2\phi_a)).
\end{equation}
Without loss of generality, we choose the local-oscillator phase to be aligned with the real part of the reflected field and therefore take $E_{\mathrm{lo}}$ to be real. Defining $P_{\mathrm{lo}}=|E_{\mathrm{lo}}|^2$, $P_r=|E_r|^2$, and $P_{\mathrm{in}}=|E_{\mathrm{in}}|^2$, the powers at the outputs of the 50:50 non-polarizing beam splitter (NPBS) are
\begin{align}
\begin{split}
P_1
&=
\frac{1}{2}|E_{\mathrm{lo}}+iE_r|^2 \\
&=
\frac{P_{\mathrm{lo}}+P_r}{2}
-2\sqrt{P_{\mathrm{lo}}P_{\mathrm{in}}}\, r_{\phi} \phi_a,
\\
P_2
&=
\frac{1}{2}|iE_{\mathrm{lo}}+E_r|^2 \\
&=
\frac{P_{\mathrm{lo}}+P_r}{2}
+2\sqrt{P_{\mathrm{lo}}P_{\mathrm{in}}}\, r_{\phi}\phi_a .
\end{split}
\end{align}
Here, we adopt the standard symmetric beam-splitter convention in which reflection introduces a factor of $i$, corresponding to a $\pi/2$ phase shift relative to transmission~\cite{GerryKnight2005}.

The total detected DC power is
\begin{equation}
P_1+P_2=P_{\mathrm{lo}}+P_r,
\end{equation}
whereas the difference signal is
\begin{equation}
P_2-P_1
=
2\sqrt{P_{\mathrm{lo}}P_{\mathrm{in}}}r_{\phi}(2\phi_a)
=
4\sqrt{P_{\mathrm{lo}}P_{\mathrm{in}}}r_{\phi}\theta_F.
\end{equation}
In the limit $P_{\mathrm{lo}}\gg P_r$, which is readily satisfied near critical coupling, the photon-shot-noise amplitude is
\begin{equation}
\mathcal{A}_{\text{psn,ho}}\approx\sqrt{2P_{\mathrm{lo}}\hbar \omega}.
\end{equation}
The corresponding homodyne SNR is
\begin{equation}
\text{SNR}_{\text{ho}}
=
2\sqrt{\frac{2P_{\mathrm{in}}}{\hbar \omega}}r_{\phi}\theta_F.
\end{equation}
The SNR gain relative to single-pass Faraday rotation is therefore
\begin{equation}
G_{\text{ho,r}}
=
\frac{\text{SNR}_{\text{ho,r}}}{\text{SNR}_{\text{sp}}}
=
2r_{\phi}.
\end{equation}
Using Eq.~\eqref{eq:genericEnhancement} and evaluating the intracavity power
enhancement on resonance at $\phi_a=0$, the reflection sensitivity
enhancement is
\begin{equation}
\mathcal{E}_{\mathrm{ho,r}}
=
\frac{2r_{\phi}}{\sqrt{\mathcal{P}_0}}.
\label{eq:SEHomo}
\end{equation}
Its dependence on finesse, internal loss, and cavity coupling is shown in Fig.~\ref{fig:homodyne}(b).

At critical coupling, Eq.~\eqref{eq:SEHomo} can be simplified to obtain the scaling with finesse. Using Eq.~\eqref{eq:r_phi} and Eq.~\eqref{eq:P0} to express $r_{\phi}$ in terms of $\mathcal{P}_0$ and $\mathcal{F}$, together with $\sqrt{\rho}\approx1$ and $\mathcal{F}\approx\pi/(1-R_1)$ in the high-finesse limit, gives
\begin{align}
\begin{split}
\mathcal{E}_{\text{ho,r}}
=
2r_{\phi}/\sqrt{\mathcal{P}_0}
&\approx
\sqrt{\mathcal{P}_0}
\\
&\approx
\sqrt{2(1-R_1)\frac{\mathcal{F}^2}{\pi^2}}
\\
&\approx
\sqrt{\frac{2\mathcal{F}}{\pi}}.
\end{split}
\end{align}

Homodyne detection in transmission uses the analogous configuration shown in Fig.~\ref{fig:homodyne}(c). Following the same argument as for reflection gives
\begin{equation}
\mathcal{E}_{\text{ho,t}}
=
\frac{2t_{\phi}}{\sqrt{\mathcal{P}_0}}.
\end{equation}
The optimized transmission enhancement and its dependence on finesse, internal loss, and coupling are shown in Fig.~\ref{fig:homodyne}(d). 

As illustrated in Fig.~\ref{fig:homodyne}(b,d), critical coupling maximizes the enhancement at fixed internal loss for both reflection and transmission.  As with side-of-fringe readout in transmission, here over-coupling, i.e., $r_0>0$ or $\tilde{r}_0>0$, gives better enhancement at constant $\mathcal{F}$. This reflects a reduced loss, i.e., a $\rho$ closer to unity. For equal $\rho$, critical coupling is again optimal. 

\subsection{Pound--Drever--Hall Detection}

\begin{figure}[h!]
    \centering
    \includegraphics[scale=0.6]{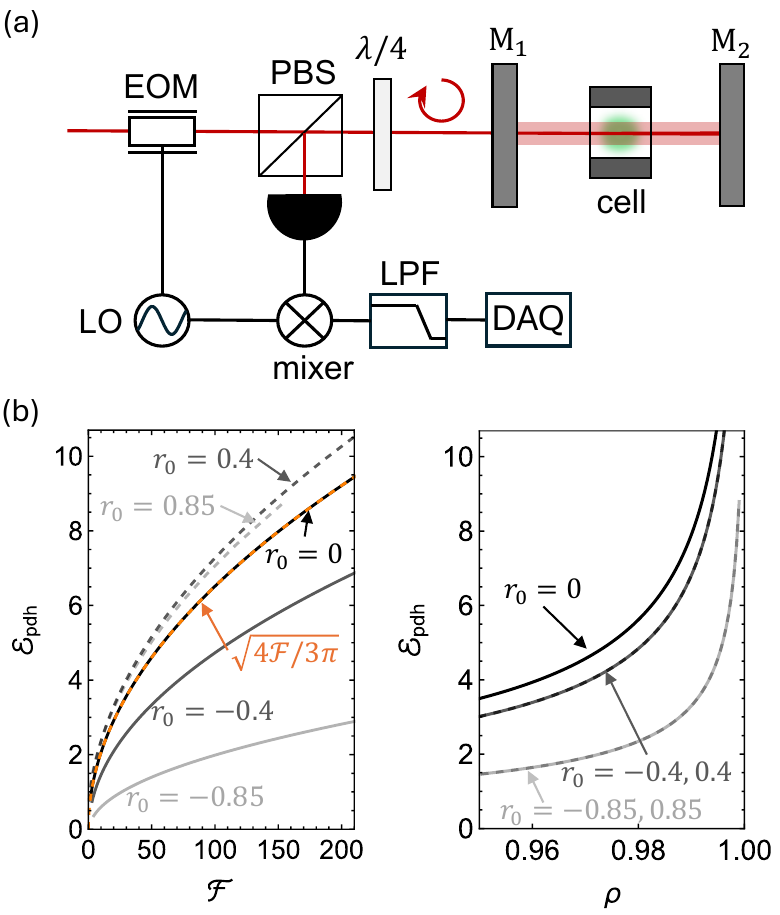}
    \caption{\textbf{Pound--Drever--Hall sensitivity enhancement.}
(a) Apparatus for PDH detection.
(b) Optimal reflection sensitivity enhancement, calculated using
Eq.~\eqref{eq:pdhEnhancement}, as a function of cavity finesse $\mathcal{F}$
(left) and intracavity amplitude-transmission factor $\rho$ (right) for
different cavity-coupling regimes. For each finesse, $r_2^2=0.999$ is fixed,
$\rho$ is varied, and $r_1$ is chosen to obtain the specified value of $r_0$.
In both plots, the carrier is resonant with the cavity ($\Delta_c=0$), the
sidebands are assumed to be fully reflected, and the modulation depth is
$\beta=1.08$. We take $L_{\mathrm{cav}}=1~\mathrm{cm}$, corresponding to a free spectral
range $\nu_{\mathrm{FSR}}\approx15~\mathrm{GHz}$.}
    \label{fig:pdh}
\end{figure}

In Pound--Drever--Hall (PDH) detection, the input probe field is phase
modulated at angular frequency $\Omega$ with modulation index $\beta$,
such that
\begin{align}
E_{\mathrm{in}}(t)
&= \sqrt{P_{\mathrm{in}}}\,
\myexp{-i\left[\omega t+\beta\sin(\Omega t)\right]} \\
&= \sqrt{P_{\mathrm{in}}}
\sum_{k=-\infty}^{\infty}
J_k(\beta)\myexp{-i(\omega+k\Omega)t},
\label{eq:probeInPDH}
\end{align}
where $J_k(\beta)$ is the Bessel function of the first kind of order
$k$. Interference between the reflected carrier and modulation
sidebands produces the PDH error signal that encodes the cavity phase
response. A representative experimental configuration is shown in Fig.~\ref{fig:pdh}(a). During the FID, the atom-induced cavity-frequency shift modulates this error signal, and the amplitude of the resulting oscillatory component defines the measured cavity-enhanced FID amplitude.

In the fast-modulation regime, $\Omega \gg \Gamma_{\mathrm{cav}}$, the
first-order sidebands lie far outside the cavity HWHM linewidth and may
therefore be treated as fully reflected. Retaining only the carrier and
first-order sidebands, the slope of the demodulated PDH error signal is
proportional to $J_0(\beta)J_1(\beta)$ and is maximized at
$\beta \approx 1.08$~\cite{BlackAJP2001}. Although higher-order
sidebands are present at this modulation depth, their amplitudes are relatively small, and their contribution to
the PDH signal is neglected in this approximation.
Assuming that the carrier is resonant with the cavity, $\Delta_c=0$, the
reflected carrier field $E_{r,0}$ and first-order sideband fields
$E_{r,-1}$ and $E_{r,+1}$ may be written as
\begin{equation}
\begin{aligned}
E_{r,0}(t)  &=
J_0(\beta) \sqrt{P_{\mathrm{in}}}
\left[r_0+i r_{\phi}(2\theta_F)\right]
\myexp{-i\omega t}, \\
E_{r,-1}(t) &=
-J_{-1}(\beta)\sqrt{P_{\mathrm{in}}}
\myexp{-i(\omega-\Omega)t}, \\
E_{r,+1}(t) &=
-J_{+1}(\beta)\sqrt{P_{\mathrm{in}}}
\myexp{-i(\omega+\Omega)t}.
\end{aligned}
\label{eq:pdh_fields}
\end{equation}
The error signal arises from interference between the reflected carrier and sidebands. The total optical power incident on the photodetector is proportional to
\begin{equation}
    P_{\rm total}(t) = 
    \big|E_{r,0}(t)+E_{r,-1}(t)+E_{r,+1}(t)\big|^2 .
\end{equation}
Retaining the DC, $\Omega$, and $2\Omega$ Fourier components gives
\begin{align}
\begin{split}
P^{\rm DC}_{\rm total}
&=
P_{\rm in}
\left[
|J_0(\beta)|^2
\left(r_0^2+4r_{\phi}^2\theta_F^2\right)
+
|J_1(\beta)|^2
+
|J_{-1}(\beta)|^2
\right], \\
P^{\Omega}_{\rm total}
&=
8P_{\rm in}
J_0(\beta)J_1(\beta)
r_{\phi}\theta_F
\sin(\Omega t), \\
P^{2\Omega}_{\rm total}
&=
-2P_{\rm in}J_{1}^2(\beta)
\cos(2\Omega t).
\end{split}
\label{eq:pdh_power_components}
\end{align}

After photodetection, the signal is mixed with the local oscillator
$V_{\rm lo}\sin(\Omega t)$ and low-pass filtered, as illustrated in
Fig.~\ref{fig:pdh}(a). The resulting baseband PDH error signal is
\begin{equation}
    \epsilon(t)
    =
    4P_{\rm in}V_{\rm lo}
    J_0(\beta)J_1(\beta)
    r_{\phi}\theta_F(t).
\label{eq:pdh_error_signal}
\end{equation}
During the FID, the time-dependent Faraday rotation $\theta_F(t)$ causes
$\epsilon(t)$ to oscillate at the Larmor frequency. If $\theta_F$ denotes the
amplitude of this oscillation, the corresponding cavity-enhanced FID amplitude,
analogous to the single-pass amplitude in Eq.~\eqref{eq:amp_sp}, is
\begin{equation}
    A_{\text{pdh}}
    =
    4P_{\rm in}V_{\rm lo}
    J_0(\beta)J_1(\beta)
    r_{\phi}\theta_F .
\end{equation}

The photon shot noise $n(t)$ incident on the photodetector is determined by the time-dependent optical power $P_{\rm total}(t)$. Because $P_{\rm total}(t)$ contains harmonics of $\Omega$, the shot noise is nonstationary, with autocorrelation function
\begin{equation}
R_{nn}(t,\tau)
=
\langle n(t)n(t-\tau)\rangle
=
2 \hbar \omega P_{\rm total}(t)\delta(\tau),
\label{eq:pdh_shot_noise_autocorrelation}
\end{equation}
where $\delta(\tau)$ is the Dirac delta function. To calculate the photon-shot-noise-limited noise floor of the demodulated error signal, we follow the treatment of nonstationary shot noise in Ref.~\cite{Niebauer1991NonstationaryShotNoise}. Let
\begin{equation}
    x(t)=n(t)V_{\rm lo}\sin(\Omega t)
\end{equation}
denote the shot-noise contribution immediately after the mixer. Its autocorrelation function is
\begin{align}
\begin{split}
R_{xx}(t,\tau)
&=
\langle x(t)x(t-\tau)\rangle \\
&=
V_{\rm lo}^2
\sin(\Omega t)\sin[\Omega(t-\tau)]
\langle n(t)n(t-\tau)\rangle \\
&=
2 \hbar\omega V_{\rm lo}^2
\sin^2(\Omega t)
P_{\rm total}(t)\delta(\tau).
\end{split}
\label{eq:pdh_mixed_noise_autocorrelation}
\end{align}
In the final line, $\delta(\tau)$ enforces $\tau=0$ in subsequent integrals.

Because the mixer is followed by a low-pass filter with bandwidth much smaller than $\Omega$, the relevant noise is obtained by averaging the autocorrelation function over one modulation period, $T_{\Omega}=2\pi/\Omega$:
\begin{align}
\begin{split}
\overline{R}_{xx}(\tau)
&=
\frac{1}{T_{\Omega}}
\int_0^{T_{\Omega}}
R_{xx}(t,\tau)\,dt \\
&=
2 \hbar \omega V_{\rm lo}^2
\delta(\tau)
\overline{
P_{\rm total}(t)\sin^2(\Omega t)
}.
\end{split}
\label{eq:pdh_time_averaged_autocorrelation}
\end{align}

For a cavity continuously locked to resonance, we evaluate the noise floor at $\theta_F=0$. The optical power incident on the photodetector then reduces to
\begin{equation}
\frac{P_{\rm total}(t)}{P_{\rm in}}
=
|J_0(\beta)|^2 r_0^2
+
2|J_1(\beta)|^2
-
2|J_1(\beta)|^2\cos(2\Omega t),
\label{eq:pdh_total_power_on_resonance}
\end{equation}
where we have used $J_{-1}(\beta)=-J_1(\beta)$. Using
\begin{equation}
    \overline{\sin^2(\Omega t)}=\frac{1}{2},
    \qquad
    \overline{\sin^2(\Omega t)\cos(2\Omega t)}
    =
    -\frac{1}{4},
\end{equation}
Eq.~\eqref{eq:pdh_time_averaged_autocorrelation} becomes
\begin{equation}
\overline{R}_{xx}(\tau)
=
\hbar \omega V_{\rm lo}^2
P_{\rm in}
\left[
|J_0(\beta)|^2 r_0^2
+
3|J_1(\beta)|^2
\right]
\delta(\tau).
\label{eq:pdh_final_autocorrelation}
\end{equation}
Therefore, by the Wiener--Khinchin theorem, the photon-shot-noise-limited amplitude spectral density of the demodulated PDH error signal is
\begin{equation}
\mathcal{A}_{\rm psn,pdh}
=
V_{\rm lo}
\sqrt{
\hbar \omega P_{\rm in}
\big[
|J_0(\beta)|^2 r_0^2
+
3|J_1(\beta)|^2
\big]
}.
\label{eq:pdh_psn_asd}
\end{equation}

The resulting PDH SNR gain relative to single-pass Faraday rotation is
\begin{equation}
G_\text{pdh}=\frac{\text{SNR}_{\text{pdh}}}{\text{SNR}_{\text{sp}}}=\frac{2\sqrt2 r_{\phi}J_0(\beta)J_1(\beta)}{\sqrt{
|J_0(\beta)|^2 r_0^2
+
3|J_1(\beta)|^2
}}
\end{equation}
To obtain the sensitivity enhancement, we account for the additional detuning required to compensate for the increased intracavity power. Because only the carrier enters the cavity, the enhancement is
\begin{equation}
    \mathcal{E}_{\text{pdh}}= \frac{G_{\text{pdh}}}{\sqrt{\mathcal{P}_0J_0^2(\beta)}}=\frac{2\sqrt2 r_{\phi}J_1(\beta)}{\sqrt{\mathcal{P}_0\big[
|J_0(\beta)|^2 r_0^2
+
3|J_1(\beta)|^2
\big]}}
\label{eq:pdhEnhancement}
\end{equation}
The optimized enhancement predicted by Eq.~\eqref{eq:pdhEnhancement} is shown in Fig.~\ref{fig:pdh}(b) as a function of finesse and internal round-trip amplitude transmission for several coupling regimes. At fixed internal loss, the enhancement is maximized at critical coupling, $r_0=0$.

In the critically coupled regime, $r_0=0$, the preceding expression simplifies to
\begin{equation}
\mathcal{E}_{\text{pdh}}^{(\text{cc})}=\sqrt{\frac{2}{3}}\frac{2r_{\phi}}{\sqrt{\mathcal{P}_0}}\approx\sqrt{\frac{4\mathcal{F}}{3\pi}}
\end{equation}
consistent with the finesse scaling shown in Fig.~\ref{fig:pdh}(b).

\subsection{Faraday-rotation readout}

We now consider Faraday-rotation readout of an FID using a linearly polarized probe transmitted through the cavity and analyzed with a balanced polarimeter. This case is more subtle than the scalar phase-readout schemes considered above. The atomic medium produces both a differential phase shift and a differential loss for the two circular polarization components, so that $t_a^+ \neq t_a^-$. Consequently, the two circular components experience different cavity resonance conditions and different effective coupling strengths. These effects reduce the simple cavity enhancement expected from a scalar phase shift and generally produce ellipticity in the transmitted field.

For an FID, the single-pass atomic phase shift is time dependent. The cavity-enhanced Faraday signal is therefore not, in general, simply proportional to the instantaneous spin projection along the probe axis. A complete treatment would require a Fisher-information analysis using the full time-dependent cavity response, including the time-dependent photon shot noise and ellipticity. Such an analysis is beyond the scope of this work. Instead, we first treat the small-phase limit, where the circular cavity modes remain effectively degenerate, and then give an approximate expression for the regime in which mode splitting and differential loss become important.
\subsubsection{Small Faraday Angle Limit}
In the limit
\begin{equation}
    \phi_a \ll \frac{t_0}{2t_\phi} \approx \frac{\pi}{2\mathcal{F}},
\end{equation}
the cavity-mode splitting and the difference in coupling between the two circular components may be neglected. Taking the probe to be resonant with the cavity, $\Delta_c=0$, the transmitted field remains approximately linearly polarized. The cavity-enhanced Faraday rotation is then
\begin{align}
\theta_{F,\mathrm{ce}}
    &= \arg(t_{\mathrm{cav}})
      = \tan^{-1}\!\left(\frac{2t_\phi \phi_a}{t_0}\right)  \nonumber \\
    &\simeq \frac{2t_\phi}{t_0}\phi_a ,
\end{align}
where $\phi_a=\theta_F$ is the single-pass Faraday rotation.

The differential power measured by the balanced polarimeter is
\begin{align}
    A_{\theta_F,\mathrm{ce}}
    &= P_{\mathrm{in}} |t_{\mathrm{cav}}(\phi_a)|^2
       \sin(2\theta_{F,\mathrm{ce}}) \nonumber \\
    &\simeq 2P_{\mathrm{in}}t_0^2 \theta_{F,\mathrm{ce}},
\end{align}
where we have assumed $\theta_{F,\mathrm{ce}}\ll1$ and
$|t_{\mathrm{cav}}|^2\simeq t_0^2$. The corresponding photon-shot-noise amplitude density is
\begin{equation}
    \mathcal{A}_{\mathrm{psn},\theta_F}
    =
    \sqrt{2P_{\mathrm{in}}t_0^2 \hbar \omega}.
\end{equation}
Comparing with the single-pass Faraday-rotation signal and photon shot noise, Eqs.~\eqref{eq:amp_sp} and \eqref{eq:psn_sp}, gives the SNR gain
\begin{equation}
    G_{\theta_F}
    =
    2t_\phi .
\end{equation}
After accounting for the increased intracavity power, the corresponding magnetic-sensitivity enhancement is
\begin{equation}
    \mathcal{E}_{\theta_F}^{\rm small}
    =
    \frac{2t_\phi}{\sqrt{\,\mathcal{P}_0}}.
    \label{eq:smallFaradayEnhancement}
\end{equation}
This expression is identical to the small-signal transmission homodyne result, as expected: in the small-rotation limit, the balanced polarimeter measures the phase quadrature difference between the two circular polarization components.
\subsubsection{Beyond Small Faraday Angle}
We next consider an approximate treatment beyond the small-$\phi_a$ limit, where the two circular components can experience appreciable cavity-mode splitting and differential intracavity loss. We evaluate the cavity response at the maximum single-pass Faraday rotation during the FID and use this response to estimate the sensitivity degradation. This approximation does not capture the full time-dependent Fisher information, but it provides a useful estimate of the loss of enhancement caused by circular-mode splitting and ellipticity.

Let
\begin{align}
    t_+ &= t_{\mathrm{cav}} \Big|_{\scriptsize\begin{array}{l}\phi_a=- \theta_F \\ \Delta_c = 0 \\\rho=\rho_{+}
    \end{array} }, \\
    t_- &= t_{\mathrm{cav}}\Big|_{\scriptsize\begin{array}{l}\phi_a= \theta_F \\ \Delta_c = 0 \\\rho=\rho_{-}
    \end{array} }
\end{align}

be the complex transmission coefficients for the two circular polarization components. Here $\rho_\pm$ denote the corresponding internal amplitude survival factors (Eq.~\eqref{eq:atomic_power_transmission}). We write their output phases as
\begin{align}
    \phi_{+,e} &= \arg(t_+), \\
    \phi_{-,e} &= \arg(t_-).
\end{align}
When $\rho_+\neq\rho_-$, the transmitted field is generally elliptical rather than purely linearly polarized.

The balanced polarimeter signal is proportional to the Stokes parameter $S_2$,
\begin{align}
    P_2-P_1
    &= 2\,\mathrm{Im}\!\left[E_-E_+^*\right] \nonumber \\
    &= P_{\mathrm{in}} |t_+||t_-|
       \sin(\phi_{-,e}-\phi_{+,e}).
\end{align}
The corresponding photon-shot-noise amplitude density is estimated from the total transmitted power,
\begin{equation}
    \mathcal{A}_{\mathrm{psn},\theta_F}
    =
    \sqrt{
    P_{\mathrm{in}}\left(|t_+|^2+|t_-|^2\right)\hbar \omega
    }.
\end{equation}
Thus the transmitted Faraday-rotation SNR is
\begin{equation}
    \mathrm{SNR}_{\theta_F,t}
    =
    \frac{
    P_{\mathrm{in}} |t_+||t_-|
    \sin(\phi_{-,e}-\phi_{+,e})
    }
    {
    \sqrt{
    P_{\mathrm{in}}\left(|t_+|^2+|t_-|^2\right)\hbar \omega
    }
    } .
\end{equation}
Relative to the single-pass Faraday-rotation SNR, this gives
\begin{equation}
    G_{\theta_F,t}
    =
    \frac{
    \sqrt{2}|t_+||t_-|
    \sin(\phi_{-,e}-\phi_{+,e})
    }
    {
    \,\sin(2\theta_F)
    \sqrt{|t_+|^2+|t_-|^2}
    } .
\end{equation}

To convert this SNR gain into a magnetic-sensitivity enhancement, we account for the average intracavity power experienced by the two circular polarization components. Let
\begin{equation}
    \mathcal{P}_\pm = \mathcal{P} \Big|_{\scriptsize\begin{array}{l}\phi_a=\mp \theta_F \\ \Delta_c = 0 \\\rho=\rho_{\pm}
    \end{array} }
\end{equation}
be the corresponding intracavity power enhancements. We then estimate the sensitivity enhancement as
\begin{equation}
\mathcal{E}_{\theta_F}
=
\frac{
2|t_+||t_-|
\sin(\phi_{-,e}-\phi_{+,e})
}
{
\,\sin(2\theta_F)
\sqrt{
\left(|t_+|^2+|t_-|^2\right)
\left(\mathcal{P}_+ + \mathcal{P}_-\right)
}
}.
\label{eq:genFaradayEnhancement}
\end{equation}
In the limit $\phi_a\rightarrow0$ and $\rho_+=\rho_-=\rho$, one has
$|t_+|=|t_-|=t_0$, $\mathcal{P}_+=\mathcal{P}_-=\mathcal{P}_0$, and
$\phi_{-,e}-\phi_{+,e}\simeq 4t_\phi\phi_a/t_0$. Equation~\eqref{eq:genFaradayEnhancement} then reduces to Eq.~\eqref{eq:smallFaradayEnhancement}, confirming consistency between the small-signal and finite-rotation estimates.

\begin{figure} [h!]
    \centering
    \includegraphics[scale=0.55]{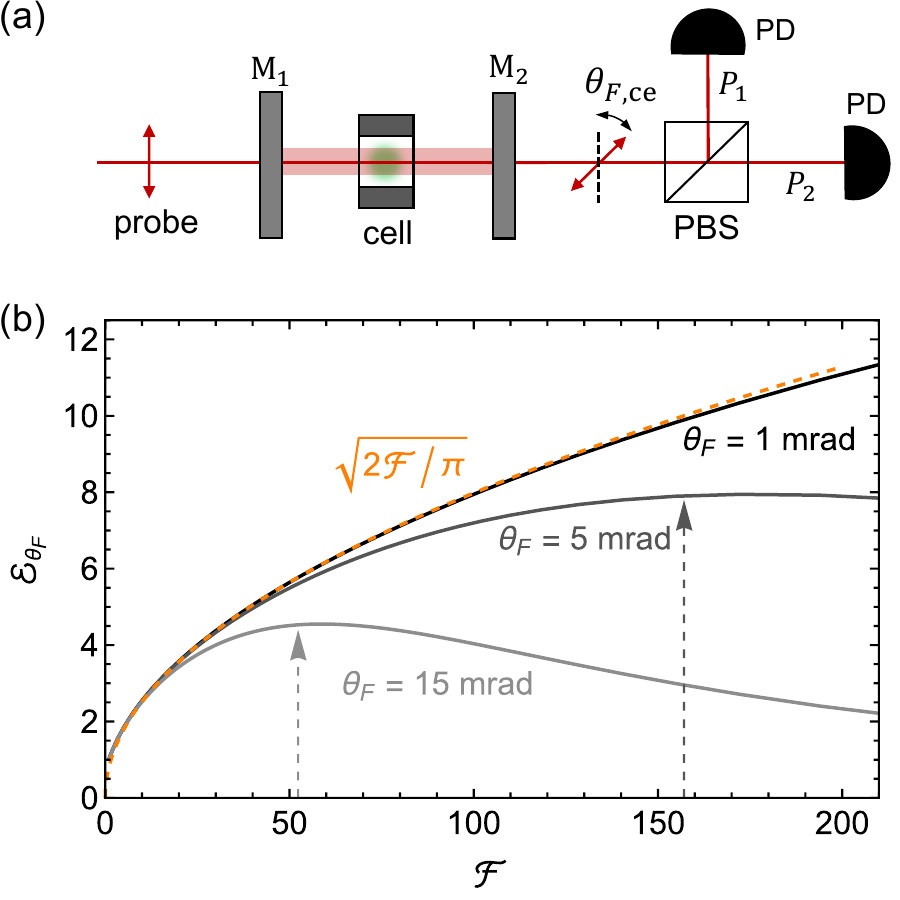}
    \caption{\textbf{Faraday-rotation sensitivity enhancement.}
(a) Apparatus for cavity-enhanced Faraday-rotation detection.
(b) Sensitivity enhancement versus cavity finesse $\mathcal{F}$ for the
indicated single-pass Faraday rotation angles $\theta_F$. The probe is
resonant with the unloaded cavity, $\Delta_c=0$. For each finesse,
$r_1^2=0.999$ is fixed, $\rho$ is varied, and $r_2$ is chosen to satisfy the
effective critical-coupling condition $\tilde{r}_0=0$. We assume a cavity
length $L_{\mathrm{cav}}=1~\mathrm{cm}$, corresponding to a free spectral range
$\nu_{\mathrm{FSR}}\approx15~\mathrm{GHz}$. The orange dashed curve denotes
the small-$\theta_F$ scaling $\sqrt{2\mathcal{F}/\pi}$. The vertical dashed lines
mark the approximate condition $\mathcal{F}\approx\pi/(4\theta_F)$ beyond
which increasing the finesse no longer improves the sensitivity enhancement.}
    \label{fig:Faraday}
\end{figure}

\section{Physical Interpretation of the Critical-Coupling Optimum}
\label{sec:CritCoupOpt}

The preceding sections show that the sensitivity enhancement is maximized at
critical coupling, or at the corresponding effective critical-coupling condition
for transmission readout. This optimum has a simple physical interpretation.
Suppose the intracavity power is held fixed while the cavity coupling is varied.
This compares the different coupling regimes at fixed light-induced atomic decoherence. The remaining
tradeoff is then between phase accumulation inside the cavity and extraction
through the detected port. An intracavity photon acquires an enhanced atomic phase shift proportional to the
effective number of passes, $2\mathcal{F}/\pi$. However, only a fraction
$t_i^2$ of the intracavity power exits through the detected port, where $t_i$ is
the corresponding amplitude transmission coefficient. Equivalently, the detected
field amplitude, and hence the photon-shot-noise-limited SNR, scales as $t_i$.
The sensitivity enhancement therefore scales schematically as
\begin{equation}
\mathcal{E}= (\text{constant}) \times
\frac{2\mathcal{F}}{\pi} t_i .
\label{eq:SchematicEnhancement}
\end{equation}
By comparing Eq.~\eqref{eq:SchematicEnhancement} with Eq.~\eqref{eq:P0}, this product is proportional to the square root
of the intracavity power enhancement at fixed input power for light input into mirror $\mathrm{M}_i$. It is
therefore maximized at effective critical coupling: $r_0=0$ for reflection readout with
$t_i=t_1$, and $\tilde r_0=0$ for transmission readout with $t_i=t_2$.

This scaling is explicit in the homodyne result in Eq.~\eqref{eq:SEHomo}. For reflection readout,
\begin{equation}
    \mathcal{E}_{\mathrm{ho,r}}
    =
    \frac{2r_{\phi}}{\sqrt{ \mathcal{P}_0}}
    \underset{\mathcal{F}\gg1}{\approx}
    \frac{t_1}{\sqrt{2}}\frac{2\mathcal{F}}{\pi}
    =
    \sqrt{\mathcal{P}_0},
    \label{eq:homoresult}
\end{equation}
which displays the $(2\mathcal{F}/\pi)t_1$ scaling of
Eq.~\eqref{eq:SchematicEnhancement}. This factor has the same coupling dependence as the intracavity field buildup,
$\sqrt{\mathcal{P}_0}$, defined in Eq.~\eqref{eq:P0} for light incident through
the detected port $M_1$. Transmission readout with $r_1>r_2$ gives the analogous result
\begin{equation}
    \mathcal{E}_{\mathrm{ho,t}}
    =
    \frac{2t_{\phi}}{\sqrt{ \mathcal{P}_0}}
    \underset{\mathcal{F}\gg1}{\approx}
    \frac{t_2}{\sqrt{2}}\frac{2\mathcal{F}}{\pi}
    =
    \sqrt{\tilde{\mathcal{P}}_0},
\end{equation}
where $\tilde{\mathcal{P}}_0$ denotes the corresponding peak intracavity power
ratio that would be obtained for light incident through mirror $\mathrm{M}_2$, introduced here only to display the coupling dependence.

Thus, in the undercoupled regime, $r_0,\tilde r_0<0$, photons remain in the
cavity long enough to acquire a large phase shift, but they are inefficiently
extracted through the detected port. In the overcoupled regime,
$r_0,\tilde r_0>0$, photons are extracted efficiently, but leave the cavity
before accumulating the full cavity-enhanced phase. Critical coupling balances
these two competing effects.

\section{Practical limits of ideal cavity enhancement}
\label{sec:cavityLimits}

This section examines two effects that can modify the idealized cavity enhancement derived in the previous section: spin-dependent absorption arising from the vector polarizability, as defined in Eq.~\eqref{eq:atomic_power_transmission}, and noise associated with the vector light shift. Spin-dependent absorption causes the cavity coupling condition itself to vary with the atomic spin state, whereas vector light-shift noise introduces an additional magnetic-field noise floor.

\subsection{Spin-Dependent Absorption}

The critical-coupling enhancement derived in the previous section assumes that
the intracavity loss is independent of the atomic spin polarization.
Spin-dependent absorption violates this assumption. As the spin polarization
precesses relative to the probe, the absorption varies over the course of the
FID, so a cavity that is critically coupled at one point in the precession
cycle becomes undercoupled or overcoupled at others. Here we estimate the
conditions under which this variation remains small enough that the cavity can
be regarded as approximately critically coupled throughout the FID.

We consider readout in reflection and take $r_2=1$ for simplicity. We further
assume the cavity is critically coupled or undercoupled, such that
$r_1\geq\rho$, and regard it as sufficiently close to critical coupling when
\begin{equation}
    |r_0|\leq\epsilon,
\end{equation}
where $\epsilon$ is a small tolerance. Combining
Eqs.~\eqref{eq:r0} and~\eqref{eq:finesse} and using $\rho\approx r_1$ near
critical coupling gives
\begin{equation}
    r_1-\rho \leq \frac{\epsilon\pi r_1\rho}{\mathcal{F}}
    \leq \frac{\epsilon\pi\sqrt{r_1\rho}}{\mathcal{F}}
    \approx \frac{\epsilon\pi r_1}{\mathcal{F}},
\end{equation}
or equivalently
\begin{equation}
    r_1\left(1-\frac{\epsilon\pi}{\mathcal{F}}\right)
    \approx r_1\exp\left(-\frac{\epsilon\pi}{\mathcal{F}}\right) \leq \rho.
    \label{eq:rhoSpinDepBound}
\end{equation}

Taking the cavity to be critically coupled at $\phi_a=|\theta_F|=0$, so that
$\rho=r_1$, Eq.~\eqref{eq:atomic_power_transmission} gives the round-trip
amplitude survival factor at the maximum spin-dependent loss as
$\rho = r_1\exp(-2\Gamma_a\phi_a/\Delta_a)$. Substituting into
Eq.~\eqref{eq:rhoSpinDepBound} yields
\begin{equation}
    \frac{2\Gamma_a}{\Delta_a}\phi_a \leq \frac{\epsilon\pi}{\mathcal{F}}.
    \label{eq:diffAbsBound}
\end{equation}
With $\phi_a$ from Eq.~\eqref{eq:phi_a}, the left-hand side scales as
$1/\Delta_a^{2}$. Optimal cavity operation
requires $\Delta_a\rightarrow\sqrt{\mathcal{P}}\,\Delta_a$ relative to the
single-pass optimum of Eq.~\eqref{eq:optimalDelta}, which suppresses the
left-hand side by $\mathcal{P}$. Since $\mathcal{P}\simeq2\mathcal{F}/\pi$ near
critical coupling, this suppression carries the same power of $\mathcal{F}$ as
the tolerance $\epsilon\pi/\mathcal{F}$. The finesse cancels, and
Eq.~\eqref{eq:diffAbsBound} reduces to a condition on the probe power alone,
\begin{equation}
    P_{\mathrm{in}} \geq \frac{
        q A_{\mathrm{eff}} \hbar\omega \Gamma_0 n_a L_a
        |\tilde{\alpha}|\langle S_z\rangle
    }{2\epsilon}.
    \label{eq:spinDepPowerBound}
\end{equation}
Spin-dependent absorption therefore imposes no fundamental restriction on the
attainable finesse, provided Eq.~\eqref{eq:spinDepPowerBound} is satisfied and
the detuning is increased accordingly. The enhancement of the accumulated
differential loss is exactly offset by the detuning increase required to hold
the light-induced relaxation rate at its optimal value. If the intracavity
power is instead compensated by reducing the input power at fixed detuning, the
left-hand side of Eq.~\eqref{eq:diffAbsBound} is unchanged and the condition
becomes an upper bound on finesse,
$\mathcal{F}\leq \pi P_{\mathrm{in}}/(2P_{\mathrm{min}})$, where
$P_{\mathrm{min}}$ denotes the right-hand side of
Eq.~\eqref{eq:spinDepPowerBound}.

As a numerical example we take the $^{87}\mathrm{Rb}$ D$_2$ line with
$\tilde{\alpha}=1$ and a fully polarized initial state, $\langle
S_z\rangle=1/2$, and parameters representative of
Ref.~\cite{HernandezPRAppl2024}, namely $L_a=1.5~\mathrm{mm}$,
$T_v=120^\circ\mathrm{C}$, $n_{\mathrm{Rb}}=2\times10^{13}~\mathrm{cm}^{-3}$,
$\Gamma_0=5000~\mathrm{s}^{-1}$, $A_{\mathrm{eff}}=0.5~\mathrm{mm}^2$, and
$q\approx4$ for high spin polarization. For a tolerance $\epsilon=0.1$,
Eq.~\eqref{eq:spinDepPowerBound} gives $P_{\mathrm{in}}\gtrsim0.2~\mathrm{mW}$,
which a typical probe power of $2~\mathrm{mW}$ exceeds by an order of
magnitude. At fixed detuning the same parameters would instead restrict the
finesse to $\mathcal{F}\lesssim16$. The power threshold rises in proportion to
the atomic column density $n_aL_a$ and the intrinsic relaxation rate
$\Gamma_0$, and falls with increasing tolerance $\epsilon$.

\subsection{Vector-Light-Shift Noise}
Vector-light-shift noise arising from photon shot noise can also limit cavity enhancement. We calculate it for a circularly polarized coherent probe, as used in all schemes considered here except Faraday rotation. A linearly polarized probe has zero mean circular polarization but comparable shot-noise fluctuations in the imbalance of its circular components, so the result also estimates the scale relevant to Faraday readout.

In the pressure-broadened regime, where the optical line is predominantly Lorentzian and its linewidth exceeds the Doppler width, the vector light shift produced by a circularly polarized probe of power $P_{\mathrm{in}}$ is equivalent to an effective magnetic field $\vec{B}_{\mathrm{LS}}=B_{\mathrm{LS}}\hat{z}$ directed along the probe axis, given by~\cite{AppeltPRA1998,Seltzer2008}
\begin{equation}
B_{\rm LS}
=
\frac{\tilde{\alpha}\sigma_0 P_{\mathrm{in}} \Gamma_a}
{2\gamma_e A_{\rm eff}\hbar\omega}
\frac{\Delta_a}{\Delta_a^2+\Gamma_a^2},
\end{equation}
where $\gamma_e = 2\pi \times 28~{\rm Hz/nT}$ is the electron gyromagnetic ratio.

Optical-power fluctuations produce corresponding fluctuations in the vector light shift. In the far-detuned limit, $\Delta_a \gg \Gamma_a$, propagating the photon-shot-noise power amplitude spectral density, $\mathcal{A}_{\mathrm{psn}}=\sqrt{2P{\mathrm{in}}\hbar\omega}$, through the power dependence of $B_{\rm LS}$ gives
\begin{equation}
\mathcal{A}_{B,{\rm LS}}
=
\left|
\frac{\partial B_{\rm LS}}{\partial P_{\mathrm{in}}}
\right|
\mathcal{A}_{\mathrm{psn}}
=
\frac{|\tilde{\alpha}|\sigma_0\sqrt{P_{\mathrm{in}}}\Gamma_a}
{\sqrt{2}\gamma_e A_{\rm eff}\sqrt{\hbar\omega}\Delta_a}.
\label{eq:vectorLSNoiseGeneral}
\end{equation}
Evaluating this expression at the optimal single-pass atomic detuning given by
Eq.~\eqref{eq:optimalDelta} yields
\begin{equation}
\mathcal{A}_{B,{\rm LS}}
=
\frac{|\tilde{\alpha}|}{\sqrt{2}\gamma_e}
\sqrt{\frac{q\sigma_0\Gamma_0}{A_{\rm eff}}},
\label{eq:vectorLSNoiseOptimal}
\end{equation}
which is independent of probe power. Equation~\eqref{eq:vectorLSNoiseOptimal}
therefore defines the vector-light-shift noise floor for the optimized
single-pass OPM considered in Sec.~\ref{sec:SinglePass}.

Introducing a cavity increases the optical power experienced by the atoms by the intracavity enhancement factor $\mathcal{P}$ defined in Eq.~\eqref{eq:intracavity_power}. If the atomic detuning and input power are held at their optimal single-pass values, the resulting intracavity power increase, $P_{\mathrm{in}}\rightarrow\mathcal{P}P_{\mathrm{in}}$, increases the vector-light-shift noise according to Eq.~\eqref{eq:vectorLSNoiseGeneral}. Optimal cavity operation, however, requires compensating for this enhancement by either increasing the atomic detuning as $\Delta_a\rightarrow\sqrt{\mathcal{P}}\Delta_a$ or reducing the input power as $P_{\mathrm{in}}\rightarrow P_{\mathrm{in}}/\mathcal{P}$, as discussed in Sec.~\ref{sec:cavityEnhanceSens}. Either adjustment exactly offsets the intracavity power enhancement in Eq.~\eqref{eq:vectorLSNoiseGeneral}. Consequently, after reoptimizing the operating conditions for the cavity, the vector-light-shift noise floor remains equal to the optimized single-pass value given by Eq.~\eqref{eq:vectorLSNoiseOptimal}.

If the background magnetic field $\vec{B}_{\mathrm{bac}}$ is perpendicular to the circular-probe axis and the magnetic field strength satisfies $B_{\mathrm{bac}}\gg B_{\mathrm{LS}}$, fluctuations in $B_{\mathrm{LS}}$ affect the measured field only at second order and are therefore negligible. To obtain a conservative upper bound, we instead consider the worst-case regime in which the vector light shift is comparable to or larger than the applied field. The maximum useful cavity enhancement is then reached when the cavity-enhanced magnetic sensitivity equals the vector-light-shift noise floor. It is therefore given by the ratio of the single-pass sensitivity $\mathcal{A}_{\mathrm{B}}$, defined in Eq.~\eqref{eq:sensitivity_sp}, to the vector-light-shift noise $\mathcal{A}_{\mathrm{B,LS}}$:
\begin{equation}
\mathcal{E}_{\rm max}
=
\frac{\mathcal{A}_B}{\mathcal{A}_{B,{\rm LS}}}
=
\frac{4\sqrt{6C_d}}{q\tilde{\alpha}^2 \langle S_z \rangle}
\frac{\gamma_e}{\gamma}
\frac{1}{{\mathrm{OD}}_{\rm sp}},
\end{equation}
where ${\rm OD}_{\rm sp}$ is the single-pass optical depth.

For a numerical estimate, we consider the Rb D$_2$ line with
$\tilde{\alpha} \approx 1$, a fully polarized initial spin state with
$\langle S_z \rangle = 1/2$, $C_d = 6$, $q\approx 4$, and $\gamma_e/\gamma \approx 4$ for
$^{87}$Rb. These values give
\begin{equation}
\mathcal{E}^{\rm max}_{^{87}{\rm Rb},D2}
=
\frac{48}{{\rm OD}_{\rm sp}}.
\end{equation}
Using parameters representative of Ref.~\cite{HernandezPRAppl2024}, namely
$L_a = 1.5~{\rm mm}$, $\Gamma_a/\pi = 23.5~{\rm GHz}$, and
$T_v = 120^\circ{\rm C}$, we find
$\mathcal{E}^{\rm max}_{^{87}{\rm Rb},D2} \approx 25$. Thus, for these
parameters, photon-shot-noise-driven vector-light-shift fluctuations do not
preclude substantial cavity enhancement. However, technical power fluctuations
above the photon shot-noise level can reduce this limit and may inhibit
sensitivity enhancement in practice.

\section{Comparison with Multipass Cells}
\label{sec:multipass}

The sensitivity enhancement of a cavity can be compared directly with that of an ideal multipass cell. The two geometries exhibit the same scaling relative to an optimized single-pass measurement, with the number of passes $M$ corresponding to the resonant intracavity power-enhancement factor $\mathcal{P}_0$. Let $M$ denote the number of passes through the atomic medium, and assume that both geometries sample the same total transverse area $A_{\mathrm{eff}}$. If the $M$ nonoverlapping beam spots in the multipass cell together fill this area, each has area $A_{\mathrm{eff}}/M$. For fixed input power $P_{\mathrm{in}}$ and atomic detuning $\Delta_a$, the local photon-scattering rate is therefore
\begin{equation}
\label{eq:gamma_sc_multipass}
\Gamma_{\mathrm{sc}}
\propto
\frac{M P_{\mathrm{in}}}
{A_{\mathrm{eff}}\Delta_a^{2}},
\end{equation}
which is $M$ times larger than in the single-pass case.

Following the argument of Sec.~\ref{sec:cavityEnhanceSens}, preserving the probe-induced relaxation rate, and hence the optimal coherence time, requires either $P_{\mathrm{in}} \rightarrow P_{\mathrm{in}}/M$ or $\Delta_a \rightarrow \sqrt{M}\Delta_a$. Since the Faraday rotation accumulated over $M$ passes gives
\begin{equation}
\mathrm{SNR}_{M}
\propto
M\frac{\sqrt{P_{\mathrm{in}}}}{\Delta_a},
\end{equation}
either constraint yields
\begin{equation}
\label{eq:snr_multipass}
\mathrm{SNR}_{M}
\propto
\sqrt{M},
\frac{\sqrt{P_{\mathrm{in}}}}{\Delta_a}.
\end{equation}
The resulting sensitivity enhancement is therefore
\begin{equation}
\mathcal{E}_{\mathrm{MP}}=\sqrt{M}.
\end{equation}

Identifying the resonant intracavity power buildup with an effective number of passes, $M_{\mathrm{eff}}=\mathcal{P}_0$, gives the same enhancement as Eq.~\eqref{eq:homoresult}. Thus, in the absence of propagation loss and for nonoverlapping beam spots sampling a fixed total area, ideal multipass and cavity geometries exhibit equivalent sensitivity scaling. Despite this equivalence, optical cavities may be better suited to miniaturized vapor cells than multipass geometries, for which realizing many well-separated trajectories within the available transverse area becomes increasingly challenging, and the smaller beam spots required can increase sensitivity to diffusion-induced coherence loss.

\section{Conclusion and Outlook}
\label{sec:conclusion}

In this work, we modeled the photon-shot-noise-limited sensitivity of cavity-enhanced OPMs in the strongly collisionally broadened regime that characterizes miniaturized vapor cells operated at high buffer-gas pressure. Using the Cram\'{e}r--Rao lower bound, we benchmarked four cavity readout schemes against an optimized single-pass FID OPM with Faraday-rotation readout. For side-of-fringe, homodyne, Pound--Drever--Hall, and Faraday-rotation readout, the optimal enhancement scales as $\alpha\sqrt{2\mathcal{F}/\pi}$, with $0.5\leq \alpha \leq 1$ a readout-dependent prefactor. At fixed internal loss, this enhancement is maximized at critical coupling, where cavity-enhanced phase accumulation is balanced against efficient extraction of the optical signal. As shown in Sec.~\ref{sec:multipass}, this result is consistent with the $\sqrt{M_{\mathrm{eff}}}$ scaling of an ideal multipass cell, whose effective number of passes is set by the intracavity power enhancement, $M_\mathrm{eff}=\mathcal{P}_0$. Near-critical coupling can be maintained throughout spin precession despite spin-dependent loss by exceeding a derived probe-power threshold and scaling the atomic detuning appropriately with finesse. Vector-light-shift noise, however, can limit the maximum useful enhancement depending on the on-resonant optical depth of the vapor cell.  

These results are particularly relevant to miniaturized and low-temperature vapor cells, where the reduced single-pass optical depth limits readout sensitivity and makes cavity enhancement especially valuable. Cavity readout may also offer practical advantages for OPMs with high on-resonance optical depth. There, large single-pass Faraday rotations can drive a balanced polarimeter out of its linear regime, increase the coupling of laser-intensity noise, and saturate the detection electronics. A suitable cavity scheme instead encodes the same atomic response as an effective resonance-frequency shift, which can be compensated or tracked through feedback to the probe frequency and thereby held near the most sensitive operating point even when the underlying atom--light interaction is strong. The dispersive interaction analyzed here also underlies quantum nondemolition measurement and conditional spin squeezing~\cite{VasilakisNPhys2015}. Incorporating measurement backaction and intracavity squeezing would extend the present analysis beyond the photon-shot-noise limit to quantify the achievable sensitivity in the spin-noise-limited regime.

Our analysis treats the probe as a single Gaussian $\mathrm{TEM}_{00}$ mode and therefore does not account for practical constraints associated with the transverse cavity-mode geometry. For millimeter-scale vapor cells, however, efficiently sampling the atomic ensemble requires a large beam waist and hence near-planar mirrors with long radii of curvature, imposing stringent fabrication tolerances. Planar Fabry--P'{e}rot cavities may therefore be the most practical geometry, although they are more susceptible to non-Gaussian mode distortion from finite-aperture diffraction and misalignment. Quantifying how these distortions limit the achievable sensitivity enhancement is an important direction for future work.

\begin{acknowledgments}
This work was supported by European Commission projects Field-SEER (ERC 101097313) and QUANTIFY (101135931); Horizon Europe (Q-Planet, 101291743), Chips Joint Undertaking (Chips JU) and the Spanish Ministry for Digital Transformation and Civil Service under reference CJU-010200-2026-6, as part of the Recovery, Transformation and Resilience Plan (PRTR),  Spanish Ministry of Science MCIN projects SAPONARIA (PID2021-123813NB-I00) and SALVIA (PID2024-158479NB-I00),  ``Severo Ochoa'' Center of Excellence CEX2024-001490-S [MICIU/AEI/10.13039/501100011033];  Generalitat de Catalunya through the CERCA program,  DURSI grant No. 2021 SGR 01453 and QSENSE (GOV/51/2022).  Fundaci\'{o} Privada Cellex; Fundaci\'{o} Mir-Puig. Views and opinions expressed are those of the authors only and do not necessarily reflect those of the European Union or the Chips Joint Undertaking. 

\end{acknowledgments}

\bibliography{Bibliography/biblio,Bibliography/MegaBib}

\end{document}